\documentclass[sigconf]{acmart}
\usepackage[utf8]{inputenc}
\usepackage{booktabs}
\usepackage{multirow}
\usepackage{graphicx}
\usepackage{subcaption}
\usepackage{amsmath}
\usepackage{float}
\usepackage{hyperref}
\usepackage{listings}
\usepackage{algorithm}
\usepackage[noend]{algpseudocode}
\usepackage{courier}
\usepackage{float}
\usepackage{color}
\usepackage[margin=10pt,font=small,labelfont=bf,labelsep=endash]{caption}
\usepackage{tikz}
\usepackage{rq2-figures/pgf-pie}
\setcopyright{none}
\renewcommand\footnotetextcopyrightpermission[1]{}

\newtheorem{definition}{Definition}[section]
\newcommand{\helium}{{\it SigGen }}
\newcommand{\codesonar}{{\it CodeSonar} }

\AtBeginDocument{%
  \providecommand\BibTeX{{%
    \normalfont B\kern-0.5em{\scshape i\kern-0.25em b}\kern-0.8em\TeX}}}

\begin{document}

\title{Reproducing Failures in Fault Signatures}

\author{Ashwin Kallingal Joshy}
\orcid{0000-0001-8236-5027}
\affiliation{
  \institution{Iowa State University}
  \city{Ames}
  \state{Iowa}
  \country{USA}
}
\email{ashwinkj@iastate.edu}

\author{Benjamin Steenhoek}
\affiliation{
  \institution{Iowa State University}
  \city{Ames}
  \state{Iowa}
  \country{USA}
}
\email{benjis@iastate.edu}

\author{Xiuyuan Guo}
\affiliation{
  \institution{Iowa State University}
  \city{Ames}
  \state{Iowa}
  \country{USA}
}
\email{xiuyuang@iastate.edu}

\author{Wei Le}
\orcid{0000-0002-6797-0648}
\affiliation{
  \institution{Iowa State University}
  \city{Ames}
  \state{Iowa}
  \country{USA}
}
\email{weile@iastate.edu}

\begin{abstract}
Software often fails in the field, however reproducing and debugging field failures is very challenging: the failure-inducing input may be missing, and the program setup can be complicated and hard to reproduce by the developers. In this paper, we propose to generate \textit{fault signatures} from the failure locations and the original source code to reproduce the faults in small executable programs. We say that a fault signature \textit{reproduces} the fault in the original program if the two failed in the same location, triggered the same error conditions after executing the same selective sequences of failure-inducing statements. A fault signature aims to contain only sufficient statements that can reproduce the faults. That way, it provides some context to inform how a fault is developed and also avoids unnecessary complexity and setups that may block fault diagnosis. To compute fault signatures from the failures, we applied a path-sensitive static analysis tool to generate a path that leads to the fault, and then applied an existing syntactic patching tool to convert the path into an executable program. Our evaluation on real-world bugs from \textit{Corebench}, \textit{BugBench}, and \textit{Manybugs} shows that fault signatures can reproduce the fault for the original programs. Because fault signatures are less complex, automatic test input generation tools generated failure-inducing inputs that could not be generated by using the entire programs. Some failure-inducing inputs can be directly transferred to the original programs. Our experimental data are publicly available at \url{https://doi.org/10.5281/zenodo.5430155}.
\end{abstract}

\maketitle

\section{Introduction}
In recent years,  software failures have caused Heathrow airport disruptions, Starbucks closing, recalls of Toyota cars, and offline of emergency calls~\footnote{\url{https://www.computerworld.com/article/3412197/top-software-failures-in-recent-history.html}}. Being able to effectively diagnose software failures is essential for bringing back the service and for avoiding future similar failures. Software development organizations like Microsoft, Google, and Mozilla deployed the failure reporting systems long  time ago and have invested huge amount of resources attempting to triage a large number of customers' crash reports received daily. However, diagnosing field failures is incredibly challenging due to the limited failure information that we can gather from the field: the program input that triggers the failure may be missing, and the environments that build and run the software can be difficult to be replicated in house where the software is developed.

In the past, the automatic support for diagnosing field failures include grouping crash reports~\cite{8449433},  computing the paths that lead to the failures~\cite{2004-FSE-Manevich}, instrumentation tools that can practically collect failure information from the field~\cite{6926857,8811942}, and the program reduction techniques that can isolate the part of the program relevant to the failures~\cite{2008-ASE-Hwa-You,2013-FSE-Sun,2015-SOSP-Kasikci,2017-IEEE-Li}. These techniques typically require stack traces, the instrumentation framework or the failure-inducing inputs.

In this paper, we present a novel and complementary solution for failure diagnosis. Our approach requires two inputs: (1) the failure location, specified as the file and the line number in the source code, and (2) the program source code. Based on the two inputs,  we compute a small and executable program that encapsulates all the important statements that can reproduce the faults.  We formally define such a program
as {\it fault signatures}.

Comparing to existing techniques of identifying a subset of statements for failure diagnosis, such as program slicing~\cite{weiser1981program, DeMillo:1996:ISSTA, Gyimothy:1999:FSE, Zhang:2006:PLDI}, and the subgraph mining on program dependency graphs~\cite{2009-ISSTA-Cheng, 2011-QRSC-Lo, 2016-QRSC-Wang}, our approach defines
a {\it fault} as a property violation and computes the property instead of dependency information to select the relevant statements, and therefore, our selection is much smaller and more precise.  Importantly, we generate an executable programs that can lead to the same fault. This enables the applications of dynamic tools, e.g.,  automatic test input generators and debuggers, for diagnosing the field failures, as opposed to only being able to inspect the code.

This paper investigated a new
concept {\it fault signature}, aiming to assist the diagnosis of field failures. We define a fault signature as an executable program that contains a minimal set of statements sufficient to produce the fault. We say a fault
signature {\it reproduces} the failure of a program when the fault signature and the program both contain the same selective sequences of statements sufficiently produce the same property violation at the same location. We show the concept of fault signatures can be applicable to a variety of faults where we are able to identify error conditions, and that fault signatures are smaller and less complex than program slicing.

This paper also presents an approach to automatically compute fault signatures. Given a failure location, we apply a path-sensitive static analysis to find the paths that lead to the failure.  We then turn the paths into an executable program.  Ideally, a path-sensitive static analysis tool is designed to identify the minimal path segments that are responsible for the faults~\cite{le2008marple, xie2007saturn, ESP}. In the past, Microsoft has used the paths reported by a static path-sensitive tool to explain the NULL-dereference errors~\cite{2004-FSE-Manevich}. Of course, static analysis is undecidable~\cite{landi1992undecidability}. When such path segments are not able to be precisely computed, our approach only reports an approximate fault signature compared to its definition.

To demonstrate the usefulness, we explored the possibility of using fault signatures to automatically generate failure-inducing inputs, and then transferring them to the original buggy software to trigger the field failures. Automatic test input generation has been an important technique for software assurance. But when software is large and complex, the tools require much computation resources and is hard to achieve the coverage needed to trigger field failures~\cite{2008-OSDI-Cadar, helin2016radamsa, zalewski2014american}. We hypothesize that by being small, focused and executable, fault signatures potentially provide a smaller search space, and demand a simpler setup that can make these tools feasible. We provided both analytical and empirical studies on the relations of fault signatures and their original programs regarding failure-inducing inputs, the root causes and the patches. This establishes the ground that, in addition to helping generate failure-inducing inputs, fault signatures may also be useful by the developers and other tools like automatic program repair. We leave the investigation and evaluation of these applications as future work.

We developed a tool, called \helium, that can compute fault signatures from real-world failures. It applied a state-of-the-art path-sensitive static analysis tool, {\it CodeSonar}, and a heuristic matching algorithm we developed to compute paths that lead to the failure specified in the given location. It then applied a syntactic patching tool, {\it Helium}, to patch the paths and generate the executable programs. In our evaluation, we successfully generated a total of 30 fault signatures that contain various types of faults from three widely used C/C++ bug benchmarks, {\it BugBench}~\cite{2005-Workshop-Lu}, {\it CoreBench}~\cite{2014-ISSTA-Bohme}
and {\it ManyBugs}~\cite{2015-IEE-Goues}. We demonstrated that all of these fault signatures are smaller, less complex and reproduced the faults in the original programs. Using two automatic test input generation
tools {\it AFL-GO}
and  {\it Radamsa}, we successfully generated failure-inducing inputs for 26 fault signatures and 8 original programs. 9 of the failure-inducing inputs generated from fault signatures are directly applicable to trigger the failures in the original programs, among which, 6 of them cannot be generated by using the original programs only. In our ablation study, we validated our hypothesis that both the simpler code logic and simpler program setup contribute to the successes of generating failure-inducing inputs from fault signatures compared to the original programs.


In this paper, we made the following contributions:
\begin{enumerate}
  \item we proposed a new concept, {\it fault signature}, for diagnosing software failures (\S 3),
  \item we developed an automatic approach that computes fault signatures and reproduces software failures given the failure location and program source code (\S 4),
  \item we analyzed the relations of fault signatures and the original programs, based on which, we applied fault signatures to automatically generate failure-inducing inputs  (\S 5), and
  \item we implemented a
        tool {\it SigGen} and empirically demonstrated that our approach is practical for real-world bugs and useful for automatically generating inputs to trigger failures in the original programs (\S 6).

\end{enumerate}

\section{A Motivating Example}
We first intuitively explain what is a fault signature and why it is useful. We used a real-world vulnerability as an example. Suppose
that {\it cvs-1.11.4} is deployed in a server, and a crash was triggered, leading to the denial-of-service (CVE-2003-0015). We know that this crash occurred in the source file server.c at line 975.

To diagnose this failure which only contains very limited information, we generated the fault signature shown in Figure~\ref{cvsex}. This executable program consists of 20 lines of code. We can directly diagnose the failure based on this program. As shown in Figure~\ref{cvsex}, {\tt free} at line~3
and {\tt xmalloc} at line~7 are always invoked together in the
function {\tt dirswitch} (called in
the {\tt while} loop at line~16), except
that {\tt dirswitch} can exit at line~5. In a loop iteration, when line~5 is invoked, the
memory {\tt dir\_name} is not allocated, and then in the next iteration, the code reaches lines~3, leading to
a {\tt double-free} crash.

\begin{figure}
	\begin{lstlisting}[commentstyle=\color{blue},
	basicstyle=\footnotesize\ttfamily,
	numbers=left
	]
void dirswitch(char* dir){
  size_t dir_len;	
  (*@\color{red}\underline{- \textbf{if} (dir\_name != NULL) free (dir\_name); }@*) 
  dir_len = strlen (dir);
  if (dir_len > 0 && dir[dir_len-1] == '/') return;
  (*@\color{olive}\underline{+ \textbf{if} (dir\_name != NULL) free (dir\_name);}@*) // patch
  dir_name=xmalloc(strlen(server_temp_dir)+dir_len+40);
  strcpy(dir_name, dir);
}

int main() {
  while(1){  
    char *cmd;
    //an example failure inducing input: 1st iteration: "Directory", 2nd: "Directory ./", 3rd: "Directory ./"
    read(0, cmd, BUFSIZE);     
    dirswitch(cmd);
  }
}
\end{lstlisting}
\caption{fault signature for cvs-1.11.4 CVE-2003-0015}
\label{cvsex}
\end{figure}


Using the fault signature in Figure~\ref{cvsex}, we successfully generated failure-inducing inputs to trigger the fault, using the test input generation
tool {\it Radamsa~\cite{helin2016radamsa}}. One example of such failure-inducing input is given at line~14 in Figure~\ref{cvsex}. In the first iteration of
the {\tt while} loop at line~12,
variable {\tt cmd} at line~15 takes the
input {\tt Directory}. When the execution reaches line~7, the memory is allocated. In the second iteration of
the {\tt while} loop, when the user provides the
input {\tt Directory ./}, {\tt free} is called for the first time at line~3
and {\tt dirswitch} exits at line~5. When the execution reaches the third iteration, {\tt free} at line~3 is executed twice, the double free is triggered and the  program crashes.

Interestingly, the failure-inducing input generated for the fault signature also triggers the double-free vulnerability in the original program. To fix the bug, we developed a patch  by
moving {\tt free}
after {\tt return} (see line~6). When we downloaded new versions
of {\it cvs}, we found that our patch developed based on the fault signature is the same as the one used
in {\it cvs-1.11.5}.

This fault is hard to diagnose in the original program. It is relevant to multiple source files and functions, as opposed to the 20 lines shown in Figure~\ref{cvsex}.  The complexities also lie in that (1) the bug is related to a global
pointer {\tt dir\_name}, which is visited by many functions, (2) the path that reproduces the fault is very long, containing a big loop with many functions invoked inside the loop, and (3) among many paths that
reach {\tt free}
in {\tt dirswitch}, only a very special path can trigger the fault. On contrast, the fault signature is small and easy to understand, as shown in Figure~\ref{cvsex}, because in the original program, only a very small number of statements along the long and complicated path actually contributed to the vulnerability. These statements demonstrate the development and context of a fault, and are useful for generating failure-inducing test inputs and patches that may be applicable in the original programs.

\section{Defining fault signatures}~\label{sec-bug-sig}
Here, we formally define fault signature and explain what it means
by {\it reproducing} a fault in the fault signature. We show that fault signature is a concept general to different types of faults and is smaller and more precise
than {\it program slice}.

\begin{figure*}
\begin{minipage}{0.5\linewidth}
\begin{lstlisting}[commentstyle=\color{blue},
	basicstyle=\footnotesize\ttfamily,
	]
void original_program(){
    ...
    char *s = (char*)malloc(10);
    char t[100] = "hello world";   
    x[20] = 'a';
    bar();
    if(strlen(t)>=10) strcpy(s,t); //fault location
    else strcat(x,t); 
    ...
}
void fault_signature(){
    char *s = (char*)malloc(10);
    char t[100] = "hello world";
    if(strlen(t)>=10) strcpy(s,t); //fault location
}
\end{lstlisting}
\end{minipage}
\hspace{0.1cm}
\begin{minipage}{0.48\linewidth}
  \includegraphics[trim=5cm 6.2cm 5cm 4cm, clip=true, width=\linewidth]{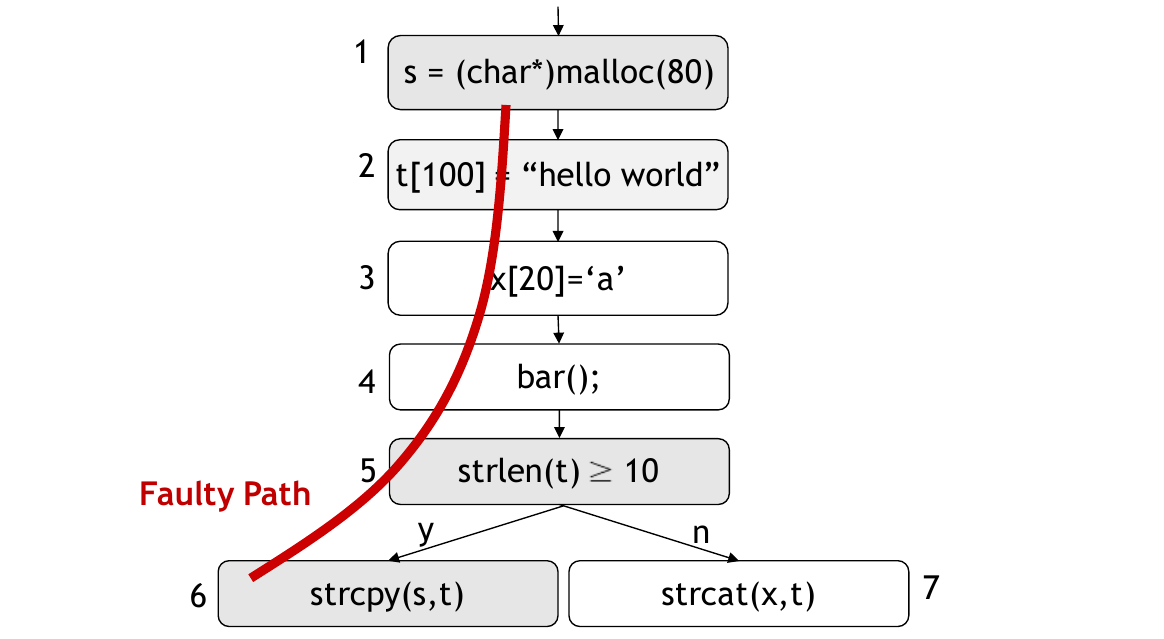}

\end{minipage}
\caption{the values of variables on the path surely lead to the fault; as long as the path segment is executed, the fault will be triggered}
\label{ex21}
\end{figure*}

\begin{figure}
\begin{minipage}{0.9\linewidth}
\begin{lstlisting}[commentstyle=\color{blue},
	basicstyle=\footnotesize\ttfamily,
	]
void original_program(){
    ...
    char* s = (char*)malloc(10);
    char t[100]; 
    read(t, STDIN_FILENO, 100); //external input
    x[20] = 'a';
    bar();
    if(strlen(x)<10) strcpy(s,t); //fault location
    else strcat(x,t);
    ...
}
void fault_signature(){
     char* s = (char*)malloc(10);   
     char t[100];   //code for compiling fault signature
     read(t, STDIN_FILENO, 100); //external input
     strcpy(s,t); //fault location
}
\end{lstlisting}
\end{minipage}
\caption{a proper external input is needed to trigger the fault}
\label{ex22}
\end{figure}

\subsection{Fault and Fault Signature}~\label{def}
Programmers make mistakes, e.g., using an incorrect variable/API or missing a check. Such a mistake will lead to incorrect conditions, also known
as {\it property violations}, in the program, and then be perceived as a failure during program execution. 

\begin{definition}
  A {\it fault} is a violation of a correct condition at a program point. We call such a program point
  the {\it fault location}.
\end{definition}

A fault consists of two key factors: the correct condition and the fault location. The negate of a correct condition is
called {\it fault condition}
or {\it error condition}.The condition and location can be derived from general safety properties or from program-specific functional bugs. For example, a buffer overflow is a fault. It violates the correct condition that the buffer size should be larger than the length of the string stored in the buffer. The fault location is the buffer access, e.g., writing to a buffer or reading from a buffer. A null-pointer dereference is a fault. It violates the constraint that a dereference should only be performed on a non-null pointer. The fault location is the pointer dereference. For program-specific bugs, we use assertions to define the correct conditions and the corresponding fault locations. As an example, if we require for all the inputs larger than 0, the output should not be 0, we
add {\tt assert(in'$>0$ \&\& out$\neq0$||in'$\leq 0$)} at the output location.

To determine if a program can violate the correct condition at the fault location, we need to identify a set of statements that are relevant to produce the conditions. Depending on whether the fault is always triggered or triggered only by certain inputs, we identified two cases and constructed two simple examples to explain
them.

  {\noindent\bf{Example 3.1:}} In Figure~\ref{ex21} on the left, {\tt original\_program} contains a fault, and the fault location
is {\tt strcpy(s,t)}. Its control flow graph (CFG) is shown on the right. We highlighted the path that leads to the fault in red. The shaded nodes represent the nodes that are relevant to produce the fault condition: when executing the sequence 1, 2, 5 and 6, we obtain a
buffer {\tt s} whose size is 10 bytes, a
string {\tt "hello world"}, and an incorrect bounds-check {\tt strlen(t)$\geq$10}. As a result,
buffer {\tt s} is overflowed at node~6, in dependent of the values before node 1 and in nodes 3 and
4.

  {\noindent\bf{Example 3.2}:} In Figure~\ref{ex22}, we
modified {\it Example 3.1} by changing the bounds-check
to {\tt strlen(x)$<10$} and added
a {\tt read} before node~2 to supply an external input. When the input size is larger than or equals to 10 bytes, the buffer overflow occurs
at {\tt strcpy(s,t)}.

The above examples demonstrated the "local property" of a fault. The faulty path segments do not have to start at the entry of the
original {\tt main} function and traverse the entire program. The beginning of the fault can be a relevant value assignment ({\tt malloc} in Example 3.1 determines the buffer size) or some external input taken in the middle of the original program (Example 3.2). According to our studies, Figure~\ref{ex22} more frequently occurred than Figure~\ref{ex21} in real-world software. For example, Figure~\ref{cvsex} is such a real-world example of Figure~\ref{ex22}.

\begin{definition}
  A {\it fault signature} of a fault $f$ is an executable program that contains only the statements sufficient to produce $f$ and to make the program executable.
\end{definition}

In Figure~\ref{ex21}, {\tt fault\_signature} contains nodes 1, 2, 5 and 6 in the CFG on the right.  The fault signature in Figure~\ref{ex22} is similar except that the bounds-check {\tt strlen(x)$<$10} is excluded as it is related to
buffer {\tt x} but
not {\tt s}.

The fault signature has two advantages: (1) it provides some context of a fault by including the path segments that lead to the fault, (2) it cuts down potentially overwhelming information by excluding the non-relevant statements. The "relevance" of a statement is determined by whether it can contribute to produce the fault condition, or in another word, whether its presence changes the condition relevant to the determination of the fault.

\begin{definition}
  We say that a fault
  signature {\it reproduces} the failure of the original program if the fault signature and the original program trigger the same fault after the two programs execute some identical subsequences~\cite{1986-Aho-Dragon} of paths.
\end{definition}

Faults are defined by a correct condition and the fault location. Two faults are
the {\it same} if they violate the same correct conditions at the same fault locations (identified by the program statements).
A  {\it subsequence} of a path means a (possibly non-continuous) sequence of statements on the path.
The {\it identical subsequences of the paths} means that the two paths, one from fault signature and the other from the original program, can be different but the subsequences obtained from the two paths respectively are identical.

During program execution, a fault in the original program is produced after executing a particular sequence of statements and arriving at the fault location, where the constraint is violated. The fault signature keeps only the statements required to produce the fault; therefore, it has an identical subsequence of the faulty path as the original program. As a result, the same fault condition is produced at the same fault location.

On the CFG in Figure~\ref{ex21}, the sequence 1, 2, 5 and 6, in the fault signature, is a subsequence of the path 1-6 in the original program. Thus, the fault signature will generate the same violation at the fault
location {\tt strcpy(s,t)} as the one in the original program.

\subsection{Fault Signatures for a Variety of Faults}\label{subsec:variety}
The fault signature concept is applicable to different types of faults, as long as we can identify the fault location and the fault condition for a fault in the program, shown in Table~\ref{fault}. The first row shows that a buffer overflow occurs when the length of the string is larger than the size of the buffer at a buffer access. The occurrence of a null-pointer dereference is always at a pointer dereference when the value of the pointer is a NULL. Double-free occurs at the C
statement {\tt free}, when the
pointer {\tt ptr} has been freed more than once, denoted
as {\tt count\_free(ptr)$>$1}, and when it is freed, the pointer is not a NULL, {\tt value(ptr)$\neq$NULL}. Resource leak violates a liveness property. The correct condition says that at the last use of the resource (fault location), the resource should be released. Thus, if along a path, the number of the allocated resources is larger than the number of the same released resources, denoted
as {\tt N\_alloc(r) $>$ N\_release(r)}, the fault occurs. For a loop, if the
loop {\it induction variable}, specified
using {\tt ind} in the table, does not change at the end of iteration, the loop will go infinite.
For functionality bugs, the programmers can use assertions to specify the correct conditions. The assertion can be about internal program states or about the input and output of the program. In this case, the fault condition is $\neg assert$.

\begin{table}[htb]
  \centering
  \caption{Different Types of Faults}~\label{fault}
  \resizebox{\columnwidth}{!}{
    \begin{tabular}{l||l|l}\hline
      {\bf Fault Type}             & {\bf Fault Location}            & {\bf Fault Condition}           \\\hline\hline
      buffer overflow              & buffer access                   & len(str) $>$ size(buf)          \\\hline
      null-pointer deref           & pointer deref                   & value(ptr) = NULL               \\\hline
      \multirow{2}{*}{double free} & \multirow{2}{*}{free a pointer} & count\_free(ptr) $>$ 1 $\wedge$ \\
                                   &                                 & value(ptr)$\neq$ NULL           \\\hline
      resource leak                & end use of resource             & N\_alloc(r) $>$ N\_release(r)   \\\hline
      infinite loops               & end of iteration                & value(ind) = value'(ind)        \\\hline

      program specific             & assertion                       & $\neg$ assert                   \\\hline

    \end{tabular}
  }
\end{table}

The six examples in Table~\ref{fault} cover a variety of types of faults. Double-free and null-pointer dereferences are the violations of the safety properties, while resource leak is the violation of the liveness property. These three are also regarding temporal properties while buffer overflow and infinite loops are related to the values and dataflow of variables. The first five types are general to all C/C++ programs, and the last one is program specific but using a general approach of assertions to define correct conditions for different program
contexts.

  {\noindent\bf{Example 3.3:}} In addition to the double-free, buffer overflow examples discussed in the previous sections, in Figure~\ref{find1}, we show that fault signature is also feasible, small and useful to capture the functional bug. This fault is
from {\it findutils 4.3.6}~\footnote{\url{https://savannah.gnu.org/bugs/?20005}}, and can lead to an incorrect output. When a user types "find tmp -mtime -2" to retrieve a file from the
directory {\tt tmp} strictly less than 2-days old, the program returns a file exactly 2-days old.

\begin{figure}
\begin{minipage}{\linewidth}
	\begin{lstlisting}[commentstyle=\color{blue}, 
	basicstyle=\footnotesize\ttfamily,
	numbers=left
	]
int main(int argc, char*argv[]){//failure-inducing input: find tmp -mtime -2
   ... 
   for (int i = end; i < argc && !looks_like_expression(argv[i], true); i++){}
  parse_time(argv, &i);
  ...
}
boolean parse_time(char* argv, int *arg_ptr){
  char* timearg; 
  struct time_val tval;
  enum comparison_type comp;
  time_t ori; 
  collect_arg(argv, arg_ptr, &timearg);
  if(get_comp_type(&timearg, &comp)){};  
  (*@\color{olive}\underline{+ \textbf{if}(comp!=COMP\_EQ) timearg--;}@*) // patch
  if(get_comp_type(&timearg, &tval->kind)){}; 
  assert(comp==tval.kind); // fault location 
  return 0;
}
int collect_arg(char** argv,int* arg_ptr,char** c_arg){
  c_arg = argv[*arg_ptr];  return 1;
}
int get_comp_type(char** str, enum comparison_type* comp_type){
  switch (**str){
    case '+':
      *comp_type = COMP_GT;
      (*str)++; break;
    case '-':
      *comp_type = COMP_LT;
      (*str)++; break;
    default: *comp_type = COMP_EQ; }
  return 1;
}
	\end{lstlisting}
\end{minipage}
\caption{fault signature of findutils-4.3.6 Savannah \#20005}
\label{find1}
\end{figure}


At line~4,
the {\tt main} function
calls {\tt parse\_time} that takes the command line
input {\tt argv}. At line~8, the
variable {\tt timearg} will store the input relevant to
the {\tt -mtime} option. For example, given "find tmp -mtime -2", {\tt timearg} will be assigned to "-2" at line~12 by the
function {\tt collect\_arg}. At line~13, {\tt get\_comp\_type} is executed and at lines 28--29, {\tt timearg} is changed to "2" and "comp" is set to "COMP\_LT". At line~15, {\tt get\_comp\_type} is called again. Because the current value
of {\tt timearg} is "2", {\tt default} at line~30 is excised. As a result,
variable {\tt tval.kind}
gets {\tt COMP\_EQ} after executing line~15, and the assertion at line~16 is triggered. This assertion says that the input time constraint, stored
in {\tt comp}, and the output time constraint, stored
in {\tt tval\_kind}, should be the same.

Based on the fault signature, we developed a patch shown at line~14. It
restores {\tt timearg} to "-2", and the assertion at line~16 is then no longer triggered. This patch can be directly transferred to fix the original program.

\subsection{Fault Signatures vs. Program Slices}
The key differences between a fault signature and a program slice include (1) fault signature is executable (2) besides the code that makes program executable, any statements in the fault signature contribute to the production of the fault condition at the fault location; while program slices use "dependency" to determine if a statement is relevant to the fault. The biggest criticism for using program slices to diagnose the fault is that they are too big. By using fault condition instead of dependency to determine the "relevance" of a statement, fault signature is expected to be smaller and more focused. Here, we constructed a simple example to further explain why a fault signature is smaller than a program slice.

\begin{figure}[ht]
  \centering
  \includegraphics[trim=5cm 3cm 5cm 2cm, clip=true, width=\linewidth]{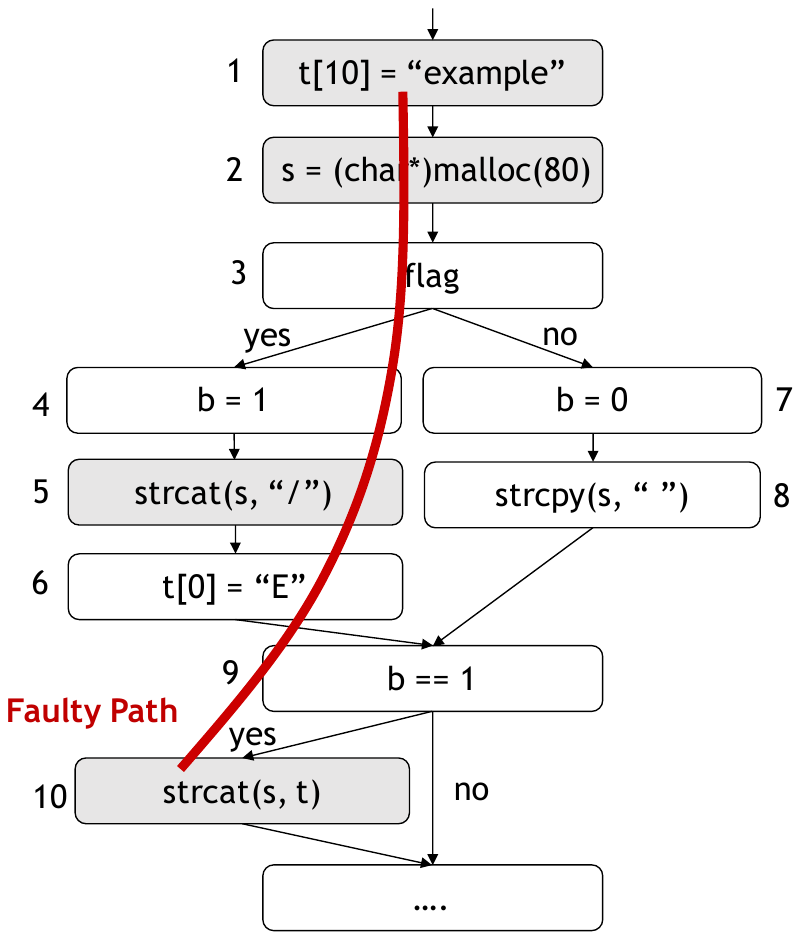}
  \caption{Comparing fault signature with program slice}
  \label{fig:ex3}
\end{figure}

In Figure~\ref{fig:ex3}, node~10 is the fault location. There are two paths traversing node 10: path $\left<1-6,9-10\right>$ contains an off-by-one fault, and another path is a safe path. The shaded boxes display a faulty path segment in the fault signature, including the nodes 1, 2, 5 and 10.

\noindent{\bf Static and dynamic slices:} Using node~10 as a slicing criterion, the static slice will include all the 10 nodes. Nodes 1,2,5,6,8 and 10 are data-dependent, as these nodes modify either
buffer {\tt s} or
string {\tt t}. Node 9 is a control-dependent node, and nodes 3, 4 and 7 are transitively included in the slice because of node 9. Dynamic slicing is also called execution slicing~\cite{HSFal405}. It computes the slice on a set of statements that have been executed. The dynamic slice based on the execution trace  $\left<1-6,9-10\right>$ will include all the nodes along the trace. That is, in addition to the shaded nodes, node 6 is included as data-dependent, and nodes 9, 4 and 3 are included as a result of control dependency analysis.

\noindent{\bf Fault signature:} When constructing the fault signature, the path $\left<1-3,7-10\right>$ is safe, and thus nodes~7 and 8 are excluded. Node 6 is a data-dependent node along the faulty path, but this statement does not contribute to the fault condition (in another word, removing this statement, the fault still occurs), and thus it does not include in the fault signature. Similarly, the control-dependent node 9 and its control and data-dependent nodes of 3, 4 and 7 do not contribute to the fault condition and thus are excluded.

This example demonstrated that program slices can include nodes along the safe paths, and the nodes that have indirect dependencies but does not contribute to the fault condition. On the other hand, fault signature provides the focus on the nodes that produce the fault conditions, removing any nodes in the fault signature, the fault condition will not occur.

\section{Computing fault signatures}
\begin{figure}

\tikzstyle{block} = [rectangle,draw,text width=6.2em, text centered, rounded 
corners, minimum height=2em]
\tikzstyle{io} = [rectangle, text width = 6.0em, align=center]
\tikzstyle{io2} = [rectangle, text width = 7.0em, align=right]
\tikzstyle{output} = [rectangle, text width = 9.0em, align=left]
\tikzstyle{bigbox} = [rectangle,draw,text width=11em, text centered, rounded 
corners, minimum height=10em]
\tikzstyle{bigbox2} = [rectangle,draw,text width=11em, text centered, rounded 
corners, minimum height=7.4em]
\tikzstyle{line} = [draw, -latex']
\resizebox{0.9\columnwidth}{!}{
\begin{tikzpicture}[auto]
	\node[io2](input){failure location, source code};
	\node[block, right of=input,node distance= 3cm](pcode){Fault analysis};
	\node[block, right of=pcode,node distance= 3.2cm](test){Path-sensitive static analysis};
	\node[block, below of=test, node distance= 3.2cm](random){Syntactic patching};
	\node[block, left of=random, node distance= 3.2cm](final){Test input generators};
	\node[io2, left of=final, node distance=3cm](rank){failure-inducing inputs};

	\draw [thick,->] (input) -- (pcode);
	\draw [thick,->](pcode) -- node[above=4pt]{fault} (test);
	\draw [thick,->](test) -- node[left=2pt]{faulty path segment} (random);
	\draw [thick,->](random) -- node[above=13pt]{fault signature}(final);
	\draw [thick,->](final) -- (rank);
\end{tikzpicture}
}

	\caption{The Workflow}
	\label{fig:flowChart}
	
\end{figure}
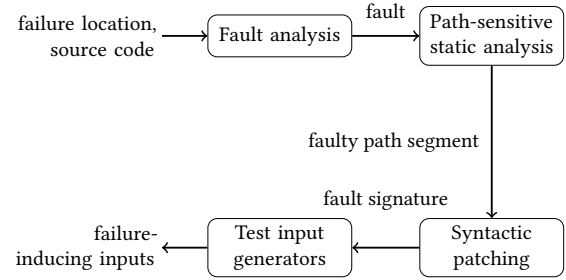

Given {\it Definition 3.2}, there can be different approaches to compute fault signatures. In this paper, we propose to apply path-sensitive static analysis to identify the faulty paths based on the failure information, and then generate executable programs. Previous research has demonstrated that paths reported by static analysis tools can provide rich information to explain how a bug is developed in static warnings and dynamic failures caused by NULL-pointer dereferences~\cite{2004-FSE-Manevich,xie2007saturn,ESP}.

Shown in Figure~\ref{fig:flowChart}, we take the failure location and program source code as input. The failure location is described as the source file and the line number where the failure occurs. We use three steps to generate a fault signature.

In the first step, \helium performs fault analysis to obtain the fault condition. It compared the statement at the failure location with a list of predefined types shown in Table~\ref{fault}. Based on the programming languages, the types listed
under {\it Fault Location} in Table~\ref{fault} are instantiated with concrete APIs and statements provided by our templates. For example,
A {\it buffer-access} can happen
at {\tt strcpy(a,b)}
and {\tt
    a[d]=e} in C/C++ programs. If any of the types is a match, the fault condition is derived correspondingly based on using the variables at the fault location.

In the second step, we provide the fault condition and the fault location to a path-sensitive static analysis tool to compute the paths that lead to the fault, which we
call {\it faulty path segments}. Ideally, such a tool can report a "complete" faulty path, traversing from where the relevant values are defined until the fault location is reached. In addition, to support developers better diagnose the static warnings, such a tool pinpoints the important statements that determine the fault conditions. In our implementation, we used a state-of-the-art commercial tool \codesonar to compute the faulty path segments.

In the third step, we
apply {\it Helium}, a syntactic patching tool to convert the faulty path segments into an executable
program. {\it Helium} generates small patches to make the code compilable and also helps resolve the dependencies to build the programs. Importantly, the patches are proved to
be {\it safe}: they do not change the order of statements on the faulty paths~\cite{kallingal2021validating}.

After fault signatures are computed, we can apply software assurance tools such as  automatic test input generators and debuggers on the fault signatures to compute diagnostic information such as failure-inducing inputs and the values at break points. The fault conditions and locations can be used as test oracles. The failure-inducing inputs will trigger the fault conditions. The correct test inputs will pass the correct conditions.

\section{Fault Signature and its Original Program: Applying Fault Signatures}
Fault signatures may be used by developers and software assurance tools for failure diagnosis.  Our hypothesis is that fault signature is smaller and simpler and thus likely to be more scalable and easier to setup. As a first step of exploring the use of fault signatures, we performed an analysis on the relations of fault signatures and the original programs, shown in Table~\ref{relation}.  The first two rows paraphrase the definition of the fault signatures: a fault signature and its original programs contain the same fault in that both programs trigger the same fault condition at the same fault location; the faulty path segments, the sub-sequence of paths, that lead to the fault, are identical in the fault signature and in the original program.  In the following sections, we provided detailed analyses regarding the relations of failure-inducing input and patch in the two programs (the third and fourth rows in the table).

\begin{table}[htb]
  \centering
  \caption{Relations of fault signature and original program }~\label{relation}
  \begin{tabular}{l||c|c|c}
    \hline
    {\bf Diagnostic info}        & \multicolumn{3}{c}{\bf Relation}                \\\hline\hline
    {Fault condition, location } & \multicolumn{3}{c}{same}                        \\\hline
    {Faulty path segments}       & \multicolumn{3}{c}{same}                        \\\hline
    {Failure-inducing input}     & same & partial & unknown                        \\\hline
    {Patch}                      & fix  & \multicolumn{2}{c}{fix with side effect} \\\hline

  \end{tabular}

\end{table}

\subsection{Failure-Inducing Input}~\label{sec:Failure-Inducing}
Knowing which inputs fail the program is very useful for debugging the failure. However, automatically generating failure-inducing inputs for real-world software is very hard. Symbolic execution tools face the challenges of scalability and loops. Fuzzing tools have difficulties to achieve the coverage and efficiency desired.

In this paper, we propose "separation of concerns" to address this challenge: (1) targeting fault condition ---- we use fault signatures to identify the input constraint that can trigger the fault along faulty path segments, and (2) reaching the "faulty region" ---- we then identify the input constraint required to reach the beginning of the faulty path segments. Based on this idea, we classify three types of relations regarding the failure-inducing inputs for the fault signature and the ones for the original program, shown in the
row {\it Failure-inducing input} in Table~\ref{relation}.

The ideal case is listed
as {\it same} in the table. It means that the failure-inducing input generated by the fault signature is also the one that can be directly used by the original program to trigger the fault. For example, in Figure~\ref{ex22}, when we provide the input
variable {\tt t} with "hello world", we can trigger the fault in both the fault signature and the original program.  In this case, there are no branches, and thus additional input constraints, needed to satisfy to reach the beginning of the fault signature. In our evaluation, we found that such cases indeed occurred in real-world code several times, e.g., in the programs
of {\it polymorph}, {\it gzip}
and {\it ncompress}
in {\it BugBench}. Typically, the fault is located in the input validation routine.

The second case is named
as {\it partial} in Table~\ref{relation}. Here, the input of the original program may contain multiple parts, and the failure-inducing input generated from the fault signature can be used as one part of the input for the original program. In Figure~\ref{extest}, we constructed such an example by adding a condition before reaching the "faulty region" in Figure~\ref{ex22}. To generate the failure-inducing input for the original program, we can run the code with input "a.exe proceed" to reach the faulty region, and then supply the failure-inducing input generated by the fault signature, "hello world", at
the {\tt read} statement to trigger the fault in the original program.

In real-world software, We found that the constraints for triggering fault condition and for reaching faulty region is often originated independently from different parts of the inputs. For example, in Figure~\ref{cvsex}, we can generate "[any], [any] ./, [any] ./" to trigger the fault at line~3 in the fault signature. The fault signature will be reached in the original program if we
invoke {\it cvs} with the input option "Directory". By combining the two, we can obtain a failure-inducing input for the original program "Directory, Directory ./, Directory ./". Similarly, in Figure~\ref{find1}, we can generate a failure-inducing
input {\tt
    find [any] [any] -n/+n} based on this fault signature. To reach the faulty region, the original program should run with the input {\tt
find [file] -mtime [any]}. We therefore can combine the two parts and generate the failure-inducing input for the original program
as {\tt
    find [file] -mtime -n/+n}.

\begin{figure}
\begin{minipage}{\linewidth}
\begin{lstlisting}[commentstyle=\color{blue},
	basicstyle=\footnotesize\ttfamily,
	]
void original_program(int argc, char* argv[]){
    ...
    if(strcmp(argv[1], "proceed")!=0) exit(0); ...
    char* s = (char*)malloc(10);
    char t[100];
    read(t,STDIN_FILENO,100); //external input
    x[20] = 'a';
    bar();
    if(strlen(x)<10) strcpy(s,t); //fault location
    else strcat(x,t);
    ...
}
void fault_signature(){
     char* s = (char*)malloc(10);       
     char t[100];
     read(t,STDIN_FILENO, 100); //external input
     strcpy(s,t); //fault location
}
\end{lstlisting}
\end{minipage}
\caption{decomposing a failure-inducing input into two parts: one for triggering the fault condition, and the other for reaching the "faulty region"(the start of the fault signature)}
\label{extest}
\end{figure}

In practice, testing often requires to achieve a desired statement coverage. These tests likely reach the faulty region, e.g., running "program proceed" in Figure~\ref{extest} reaches the beginning of the fault signatures. The challenge is to find a special input that can follow a particular faulty path from there, and trigger the fault condition. In this case, we can use fault signatures to generate failure-inducing inputs targeting fault conditions. For example, in Figure~\ref{extest}, we generate "hello world" as an input
for {\tt read}, and then combine it with existing test "program proceed" to generate the failure-inducing input for the original programs.

We performed an empirical study on 30 fault signatures generated (see Section 6) from a set of C/C++ real-world programs. 90\% of the original programs shipped with test suites. After running the tests, we found that among these programs with test suites,  more than 75\% program have tests that can reach the "faulty region", i.e., the start of the fault signatures. There are often more than one test can reach there. These tests can be used with the failure-inducing inputs generated from the fault signatures to produce the failure-inducing inputs for the original programs.

There is also a case where the failure-inducing input generated by fault signatures cannot be directly used by the original program. One situation is that the constraint for triggering the faults and for reaching the beginning of the fault signatures are jointly posed on the same part of the inputs. The other is that the relations are hard to obtain due to the approximations applied for computing fault signatures in practice, due to e.g., pointers and library calls whose source code is unknown. We call this category
as {\it unknown} in Table~\ref{relation} and will perform further investigations in our future work.

\subsection{Patches}
To fix the fault, the patches in both the fault signature and the  original program need to modify the values related to the fault condition.  Depending on how useful a fault signature patch is, we classify two cases: (1) ideally, the patches based on fault signature can be directly transferable to the original program, listed
as {\it fix} in the
row {\it Patch} in Table~\ref{relation}, or (2) in a less ideal but still possibly useful case, the patch based on fault signature fixes the fault in the original program but leave undesirable side effects, listed
as {\it fix with side effect} in the table. We provide some examples below.

In Figure~\ref{ex21}, the root cause of the original program is the incorrect bounds-check at node~5 in the CFG, and thus the patch fixed the bounds-check shown in Figure~\ref{ex23}. This patch also can correctly fix the fault in the original program. The examples in Figures~\ref{cvsex} and ~\ref{find1} also belong to this type. On the other hand, in Figure~\ref{ex22}, when only diagnosing the fault signature, we may conclude that the code misses a bounds-check. Transferring this
patch {\tt if(strlen(x)<10)} to the original program can fix the bug. But the correct fix in the original program should also
remove {\tt if(strlen(t)<10)}. See Figure~\ref{ex24}.

In our study on real-world bugs reported in Section 6, we found that many patches modify the faulty path segments common in the fault signatures and the original programs, and thus for many faults, it may be feasible to find bug fixes based on fault signatures.
When transferring patches made from fault signatures, additional tests can be performed to avoid side effects in the original program.

\begin{figure}
\noindent\begin{minipage}{\linewidth}
	\begin{lstlisting}[commentstyle=\color{blue}, 
	emph={timearg,get_comp_type,get_relative_timestamp},emphstyle=\color{blue},
	basicstyle=\footnotesize\ttfamily
	]  
void fault_signature(){
    char* s = (char*)malloc(80);       
    char t[100] = "hello world";
 -  if(strlen(t)>=10) strcpy(s,t); //root cause
 +  if(strlen(t)<10) strcpy(s,t); //fix
}
	\end{lstlisting}
\end{minipage}
\caption{patches for Figure~\ref{ex21}: we can directly transfer this patch to the original program}
\label{ex23}
\end{figure}
\begin{figure}
\noindent\begin{minipage}{\linewidth}
	\begin{lstlisting}[commentstyle=\color{blue}, 
	emph={timearg,get_comp_type,get_relative_timestamp},emphstyle=\color{blue},
	basicstyle=\footnotesize\ttfamily
	]  
vvoid original_program(){
    ...
    char* s = (char*)malloc(10);
    char t[100];
    read(t,STDIN_FILENO,100); 
    x[20] = 'a';
    bar();
-   if(strlen(x)<10) strcpy(s,t); //root cause: used a wrong variable
+   if(strlen(t)<10) strcpy(s,t); //fix
    else strcat(x,t);
    ...
}
void fault_signature(){
    char* s = (char*)malloc(10);       
    char t[100];
    read(t,STDIN_FILENO,100); 
+   if(strlen(t)<10) //root cause: miss a bounds-check
    strcpy(s,t); 
}
	\end{lstlisting}
\end{minipage}
\caption{patches for Figure~\ref{ex22}: fault signature patch fixed the bug, but leave side effect in the original program}
\label{ex24}
\end{figure}

\section{Evaluation}
Our evaluation investigates the following research questions:
\begin{itemize}
  \item RQ1: Can our approach compute fault signatures which are smaller and less complex than real-world programs?
  \item RQ2: Can fault signatures reproduce the faults in the original programs?
  \item RQ3: Is it easier to generate failure-inducing inputs for fault signatures compared to the original programs? Can failure-inducing inputs generated for fault signatures be useful for original programs?
\end{itemize}

\subsection{Experimental setup}~\label{subsec:ExpSetup}
{\it Implementation:} We used \codesonar to compute faulty paths segments
and {\it Helium} for syntactic patching to convert paths to executable. We used the scripts to automate this process from the given failure location and source code. For functional bugs where assertions are not readily available, we semi-automatically generated assertions by comparing the invariants (running Diakon~\cite{1999-ICSE-Ernst} with provided tests) of correct and buggy versions, and by analyzing the patches provided by the benchmark.  We validated the assertions by executing the test provided with the benchmarks to ensure that failure-inducing inputs trigger the assertions and that correct inputs pass the
tests.

  {\it Software subjects:} We used three popular  C/C++ bug benchmarks,  {\it BugBench}~\cite{lu2005bugbench}, \textit{CoreBench}~\cite{2014-ISSTA-Bohme} and \textit{ManyBugs}~\cite{2015-IEE-Goues}. These benchmarks contain a total of 20 real-world buggy programs. The three benchmarks cover a variety of types of faults, including buffer overflows, null-pointer dereferences, double-free, resource leaks, and program specific functional bugs. We used the failure-inducing input provided by the benchmark to create the
failures.

  {\it Experimental design:} For RQ1, we report the total number of fault signatures computed using our approach. We present the evidences that fault signatures are smaller and less complex using metrics of code size (LOC)
and {\it Cyclone Complexity}~\cite{1976-IEE-McCabe}. We compared fault signatures with the original programs, the unit tests (invoking all the functions the fault signatures traverse), and the program slices  (we used the
tool {\it Atlas}~\footnote{\url{www.ensoftcorp.com/atlas/}}).

For RQ2, to study if a fault signature contains the same fault as the original program, we ran failure-inducing input on the original program, and when reaching the beginning of the fault signature, we switch the execution to the fault signature and observe (
using {\it Valgrind}~\footnote{\url{https://valgrind.org/}}) whether the execution triggers the same failure (same fault condition at the fault location) in the original program. We also manually generated at least one test input for each fault signature to see if the same failure is reproduced. In addition, we automatically transferred developer's patches (documented in the benchmarks) which fix the bug in the benchmark programs to the fault signature. We built the patched fault signatures and ran fuzzing with Radamsa for 2 hours to validate the patches.

For RQ3, we used AFL-GO (a greybox fuzzer) and Radamsa (a black box fuzzer) to automatically generate failure-inducing inputs for both fault signatures and the original programs. We studied the fuzzing tools available~\cite{DBLP:journals/corr/abs-1812-00140}, and selected two that are well-maintained, state-of-the-art and can work with real-world C/C++ programs. We used the fault condition and fault location as test oracles, and
used {\it Valgrind} to determine if the fault is triggered.

The initial seeds for fault signatures were randomly generated based on the inputs and their types. For the original programs, we use a random test from the program's own test suite, or non-faulty input provided with the bug benchmarks when test suites are not available. Both the fault signature and original programs were fuzzed for one hour using Radamsa and AFL-GO. We used Radamsa's \texttt{-seed} flag to ensure reproducibility and used the default values from OSS-FUZZ~\footnote{\url{https://github.com/google/oss-fuzz}} as the parameters for AFL-GO.

All of our experiments were run on a VM with an 8 core Intel Haswell processor, 16 GB memory and CentOS 8.

\subsection{Results for RQ1}
Our approach computed a total of 30 fault signatures, covering the types of buffer overflows, null-pointer dereferences, double free, resource leaks and functional bugs. As shown in Table~\ref{size}, the fault signature is small with an average size of 84.46 sloc, compared to 265K sloc for the original programs (under $ori$),  155.9 sloc for unit tests (under $unit$), 10 k sloc for static slices (under $ss$), and 260.50 sloc for dynamic slices (under $ds$). Shown in the
row {\it complexity}, the fault signature is less complex in terms of the control flow of branches and paths captured by the Cyclone complexity. Such complexity is an important factor that affects the scalability of code understanding and of automatic tools such as symbolic executors.

It should be noted that the numbers reported in Table~\ref{size}  are in favor of baselines (
Columns {\it ss} {\it ds}
and {\it ori}) in that the computation for slices and Cyclone complexity only succeeded for a set of smaller programs, although we used the most mature tools we can find to compute them. Specially, 14 out of 30 programs are not able to report correct static slicing results. 10 large programs like PHP, hung after 24 hours, and for 6 programs, slicing procedures stopped at some internal functions instead of the beginning of the program. Similarly, we also failed to compute Cyclone complexity for PHP.

Although our approach computed fault signatures from large projects with different types, we observed three challenges: first, for buffer overflow,  the crash does not always immediately occur at the buffer access, as specified in Table~\ref{fault}. Second, not all the functional bugs have generated assertions. Third, static analysis is not always able to report paths given fault conditions.
This points out the future work on how to improve the effectiveness of computing fault signatures.

\begin{table}[htb]
  \centering
  \caption{Fault signatures are smaller and simpler }~\label{size}
  \begin{tabular}{l||c|c|c|c|c}\hline
    metric     & fs    & ori        & unit   & ss        & ds     \\\hline\hline
    size (loc) & 84.46 & 265,881.73 & 155.90 & 10,014.41 & 260.50 \\\hline
    complexity & 19.00 & 15,006.79  & 42.03  & -         & -      \\\hline
  \end{tabular}
\end{table}


\noindent \fbox{
  \parbox{0.95\linewidth}{\textbf{Summary for RQ1:} our approach computes 30 fault signatures of a variety of faults from real-world software. The fault signatures are smaller, simpler than original programs, unit tests, and static and dynamic slices.
  }
}

\subsection{Results for RQ2}~\label{subsec:Eval-RQ2}
\begin{figure}[h]
\hspace*{-.82in}
\begin{minipage}{.38\linewidth}
    \resizebox{\linewidth}{!}{%
    \begin{tikzpicture}[font=\HUGE]
        \pie[rotate=80,sum=auto, color = {violet!60!white,red!50!white}]
        {
        3/Not triggered,
        27/Triggered
        }
        \end{tikzpicture}
    }
    \subcaption{Testing}
    \label{fig10:testing}
\end{minipage}
\begin{minipage}{.38\linewidth}
    \resizebox{1.62\linewidth}{!}{%
    \begin{tikzpicture}[font=\HUGE]
        \pie[rotate=50,sum=auto, color = {gray,violet!60!white,red!50!white}]
        {
        3/No developer patch,
        4/Could not patch,
        23/Patched successfully
        }
        \end{tikzpicture}
    }
    \subcaption{Patch transfer}
    \label{fig10:patch-transfer}
\end{minipage}
\caption{Fault signatures reproduce the original fault}
\end{figure}
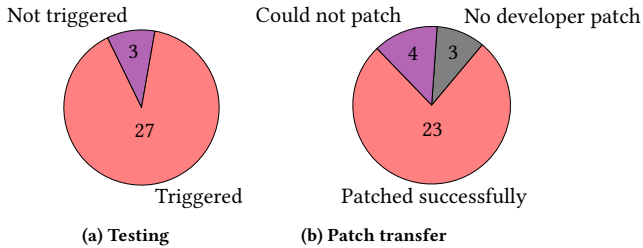

Using manually generated test inputs, we triggered the same faults for all the 30 fault signatures. When testing the original programs substituted with fault signatures using failure-inducing inputs from the original programs, we triggered 27 faults, shown in Figure \ref{fig10:testing}. We failed to trigger 3 faults.
In these cases, the static analysis tool detected the same fault, but the warning path was not the same as the path executed by the failing tests.

Among the 30 faults, 27 shipped with developers' patch. As shown in Figure~\ref{fig10:patch-transfer}, we transferred the patch successfully for 23 fault signatures. We validated these patches and found that none could trigger the original fault after fuzzing. We could not patch the 4 fault signatures because they didn't contain code that was modified by the patch.

\vspace*{.25em}

\noindent \fbox{
  \parbox{0.95\linewidth}{\textbf{Summary for RQ2:} 
    We triggered the same faults for 30 fault signatures via manual testing and for 27 fault signatures when substituting them in the original programs. Using the original patches available for 27 faults, 23 (85.18\%) fault signatures were fixed, showing that in many cases, we can modify the faulty path segments present in fault signatures to fix the faults.
  }
}

\subsection{Results for RQ3}
\begin{table}[htb]
  \centering
  \caption{Fault signatures simplify failure-inducing input generation}~\label{tbl:RQ3-input-gen}
  \begin{tabular}{l||c|c|c}\hline
                     & AFL-GO & Radamsa & Total \\\hline\hline
    Fault signature  & 11     & 24      & 26    \\\hline
    Original program & 5      & 6       & 8     \\\hline
  \end{tabular}
\end{table}

Table~\ref{tbl:RQ3-input-gen} shows that we successfully generated failure-inducing inputs for 26 out of 30 faults using fault signatures, while only generated 8 using the original programs.  For fault signatures, 9 faults are triggered by both tools, compared to 3 for the benchmark programs.

We also found that the 8 faults triggered using the benchmark programs also triggered by fuzzing fault signatures. However, the fault signatures generated the failure-inducing inputs for additional 18 faults. After further analyzing our results, we found that the input generation indeed benefits from the reduced complexity and simpler setup enabled by the fault signatures. For example, \textit{Lighttpd} required a client-server setup which couldn't be done easily on AFL-GO.  Similarly, when testing \textit{Python}, Radamsa started interactive shell that slowed down the input generation. In other programs like \textit{find} and \textit{make}, AFL-GO got stuck or slowed down when traversing complicated branches, especially the ones nested in the loops. Fault signatures do not contain such branches as they are not relevant to the faults.

\begin{table}[htb]
  \centering
  \caption{Inputs for fault signatures are useful for triggering original programs}~\label{tbl:RQ3-input-rel}
  \begin{tabular}{c||c|c|c}\hline
    Total faults & same & partial & unknown \\\hline\hline
    30           & 3    & 8       & 19      \\\hline
  \end{tabular}
\end{table}

We analyzed the failure-inducing inputs generated for fault signatures (both manually and automatically). We also analyze the code in fault signatures and original programs. As shown in Table~\ref{tbl:RQ3-input-rel}, We found among the 30 fault signatures, 3 fault signatures report the same failure-inducing inputs with the original programs, 8 are partial inputs of the original programs, and 19 are unknown. For the 11 directly usable cases, we only successfully generated the failure-inducing inputs for 3 original programs, but generated for 9 fault signatures. In another word, using fault signatures, we can generate failure-inducing inputs for 6 new faults.

\noindent \fbox{
  \parbox{0.95\linewidth}{\textbf{Summary for RQ3:}  For the 30 faults studied, we automatically generated failure-inducing inputs for 8 original programs and for 26 fault signatures. We observed the benefit of smaller code and simpler setups brought by the fault signatures. We triggered 6 more faults in the original programs using fault signatures, which cannot be triggered by running original programs alone.
  }
}

\section{Threats and Limitations}

\subsection{Threats to Validity}
{\it Internal validity}: Our approaches may compute an approximation of the fault signatures as we use paths computed by static analysis. We have validated fault signatures by running manually and automatically generated test inputs and also by substituting the fault signatures in the original programs during testing to make sure the same faults are triggered. We used semi-automatic approaches to generate assertions for functional bugs. This process may introduce biases. We used tests from the original programs to make sure our assertions are triggered only under failure-inducing inputs, and also our assertions are not triggered after patches are
applied.

  {\it External validity}: We have used real-world code of a variety of faults from three C/C++ bug benchmarks we can find. But these bugs still did not cover all the types of faults specified in Table~\ref{fault}, and these bug benchmarks may not represent all the C/C++ software.

\subsection{Limitations and Assumptions}
\label{limitations-and-assumptions}
We used the failure locations to compute fault signatures. For some faults like buffer overflow, failures may not manifest immediately at the fault locations. We plan to investigate in future work on how to determine fault locations based on certain buffer overflow failures.

Although all of our 30 fault signatures are triggered in testing, static analysis may not always be able to produce a precise path. We have observed the cases where the path contains additional regarding the fault conditions. Thus the fault signature computed this way is likely to be an approximation of its definition, although still useful as shown in our results. Static analysis also has a problem of false positives, e.g., due to infeasible paths. However, when applied to reproduce failures, we already know a failure has happened, and thus the path computed to match the failure is mostly feasible. 

Fault signatures work for a variety of types of faults. For functional bugs, we used assertions as oracles. The similar approach is used by Angelix~\cite{angelix}, which fixes the functional bugs using assertions as oracles.

The idea of fault signatures is to separate the concerns of fault production and reaching to the fault regions, and to provide automation for understanding/triggering fault conditions. In this paper, we showed that fault signatures provide the failure-inducing inputs and the simple and easy-to-run executables for debugging, which is a very hard problem. However, we need to conduct further studies to conclude if fault signatures are also useful for human fault diagnosis.

\section{Related Work}
The concept of bug and fault signatures have been proposed, but have different definitions. Hsu et al. developed RAPID~\cite{2008-ASE-Hwa-You} that used string matching to look for common segments in failed execution traces to generate a signature. Sun et al. proposed predicate bug signature mining~\cite{2013-FSE-Sun} using both data and control flow information for a signature that included statements that caused or effected the bug. Similarly, Kasikci et al. proposed failure sketching~\cite{2015-SOSP-Kasikci} for a signature that only provided statements that lead to the fault and Li et al. proposed fault-localization based minimum debugging frontier set, a model for detecting faults in execution~\cite{2017-IEEE-Li}.

Program slicing is an important technique for identifying the context of the
fault. {Researchers have worked on minimizing the slice by either intersecting static slices with run time information~\cite{article,5562955,Prez2019AutomaticTO} or  ranking the statements within the slice~\cite{6100114,HSFal405}. The representative work include model-based intersection of dynamic slice~\cite{5562955} and observation-based slicing~\cite{Prez2019AutomaticTO}. For ranking suspicious statements, dynamic slicing with change-impact analysis~\cite{6100114} and hybrid slicing that improves this further by accounting for passing test cases~\cite{HSFal405} have been used.}

Fault localization techniques aim to pinpoint the location and context of the faults. Spectrum-based localization ranks statement(s) as likely root cause based test cases and their corresponding code coverage. Initially they started by ranking a single statement~\cite{Jones:2002:ICSE, Jones:2005:ASE,2016-IEEE-Wong} and was later extended to include a set of statements~\cite{2009-Journal-Abreu,2006-PRDC-Abreu,2002-ICES-Jones,2014-IEEE-Wong,8449433} as likely root cause. Graph-based techniques include, computing discriminative sub-graphs from passing and failing traces~\cite{2009-ISSTA-Cheng,2011-QRSC-Lo},  using execution traces and nearest neighbor technique~\cite{2003-IEEE-Renieres},  and incorporating suspicious components from bug report as weighted control flow sub-graph, to produce set of statements related to the bug. Instrumentation-based approaches either selectively instrumented important statements~\cite{2014-ISSTA-Zuo} or calculated failed execution blocks from execution traces~\cite{2020-SOFL-Pan} to narrow down the context.

Compared to the above related work which all aim to identify code segments as relevant to the bug, our approach generates small executable programs that contain the same fault as the original program, which can leverage existing tools for debugging.

Our work is also related to reproduce software failures. Mathieu et al. used crash traces with model checking to reproduce failures~\cite{7081820}. Ning Chen et al. used crash stacks in combination with backward symbolic execution and method sequence composition to automatically generate test cases that reproduce the faults\cite{6926857}. Yu et al. used natural language processing with dynamic GUI even synthesis to automatically recreate crashes from Android apps~\cite{8811942}. A state-of-the-art failure reproduction tool, EvoCrash, by Monzhan et al.~\cite{7985663}, builds upon EvoSuite and uses guided genetic algorithm to search the crash stack to narrow down the faulty location and generate test cases that can reproduce it. Our approach uses source code and fault location to reproduce the fault in fault signatures.

\section{Conclusions and Future Work}
In this paper, we define a fault signature as a small, executable program that can lead to the same fault as the original program. The key idea is to decompose the complexity of fault diagnosis into two parts: (1) the form and development of the fault condition along a path, and (2) how program execution reaches the faulty region. Focusing on the only fault production (part (1)), we can create a smaller, less complex and easier-to-setup program which is useful. Our evaluation on real-world bugs and C/C++ programs showed that our approach can generate fault signatures for a variety of types of faults, using only failure location and program source code. Because it is less complex and easier to run, we generated failure-inducing inputs much more successfully than using the original programs. Among the 30 fault signatures, we obtained failure-inducing inputs that triggered 6 faults in the original programs that cannot be generated without fault signatures. In future work, we plan to further explore the applications of fault signatures and the approaches of more precisely computing fault signatures.

\section{Significance of Our Contributions}

 {\bf Impact for the industry:} Reproducing and debugging software field failures is an important but very hard problem, manifesting as the challenges of processing millions of crash reports collected by Google, Microsoft and Mozilla. Our work used minimum failure information and automatically generated failure-inducing inputs that can trigger failures in the original programs. Such failure-inducing inputs cannot be generated using original programs alone. This brings us confidence that fault signatures may also serve as a bridge to make other tools, e.g., symbolic execution and automatic repair, work better with large complex software, which we will explore in our future
studies.

  {\bf Impact for the program analysis knowledge:}  Fault signature is a formally defined new concept contributed to the program analysis literature. It is relevant to program reduction and program slicing. Those concepts have generated many works on  fault diagnosis and debugging. We hope that fault signatures can also be such an important
concept.

  {\bf Impact for the future software engineering research:} This paper defined fault signatures and showed that it is a very promising concept. We provided a new direction of fault diagnosis, i.e., generating a small relevant program for debugging and then transferring the diagnostic product to the original programs. This approach is different from program slicing, delta-debugging and spectrum-based fault location. During our studies, we have identified the need of more techniques to compute fault signatures, and many opportunities of combining with other tools, such as repair tools, symbolic executions, for applications. In the past, small programs~\cite{zitser2004testing} have been manually constructed as bug benchmarks, our work can automatically produce similar bug benchmarks.

\appendix

\bibliographystyle{ACM-Reference-Format}
\bibliography{helium}


\begin{thebibliography}{50}


\ifx \showCODEN    \undefined \def \showCODEN     #1{\unskip}     \fi
\ifx \showDOI      \undefined \def \showDOI       #1{#1}\fi
\ifx \showISBNx    \undefined \def \showISBNx     #1{\unskip}     \fi
\ifx \showISBNxiii \undefined \def \showISBNxiii  #1{\unskip}     \fi
\ifx \showISSN     \undefined \def \showISSN      #1{\unskip}     \fi
\ifx \showLCCN     \undefined \def \showLCCN      #1{\unskip}     \fi
\ifx \shownote     \undefined \def \shownote      #1{#1}          \fi
\ifx \showarticletitle \undefined \def \showarticletitle #1{#1}   \fi
\ifx \showURL      \undefined \def \showURL       {\relax}        \fi
\providecommand\bibfield[2]{#2}
\providecommand\bibinfo[2]{#2}
\providecommand\natexlab[1]{#1}
\providecommand\showeprint[2][]{arXiv:#2}

\bibitem[{Abreu} et~al\mbox{.}(2006)]%
        {2006-PRDC-Abreu}
\bibfield{author}{\bibinfo{person}{R. {Abreu}}, \bibinfo{person}{P. {Zoeteweij}}, {and} \bibinfo{person}{A.~J. c. {Van Gemund}}.} \bibinfo{year}{2006}\natexlab{}.
\newblock \showarticletitle{An Evaluation of Similarity Coefficients for Software Fault Localization}. In \bibinfo{booktitle}{\emph{2006 12th Pacific Rim International Symposium on Dependable Computing (PRDC'06)}}. \bibinfo{pages}{39--46}.
\newblock


\bibitem[Abreu et~al\mbox{.}(2009)]%
        {2009-Journal-Abreu}
\bibfield{author}{\bibinfo{person}{Rui Abreu}, \bibinfo{person}{Peter Zoeteweij}, \bibinfo{person}{Rob Golsteijn}, {and} \bibinfo{person}{Arjan J.~C. van Gemund}.} \bibinfo{year}{2009}\natexlab{}.
\newblock \showarticletitle{A Practical Evaluation of Spectrum-Based Fault Localization}.
\newblock \bibinfo{journal}{\emph{J. Syst. Softw.}} \bibinfo{volume}{82}, \bibinfo{number}{11} (\bibinfo{date}{Nov.} \bibinfo{year}{2009}), \bibinfo{pages}{1780–1792}.
\newblock
\showISSN{0164-1212}
\urldef\tempurl%
\url{https://doi.org/10.1016/j.jss.2009.06.035}
\showDOI{\tempurl}


\bibitem[Aho et~al\mbox{.}(1986)]%
        {1986-Aho-Dragon}
\bibfield{author}{\bibinfo{person}{Alfred~V Aho}, \bibinfo{person}{Ravi Sethi}, {and} \bibinfo{person}{Jeffrey~D Ullman}.} \bibinfo{year}{1986}\natexlab{}.
\newblock \showarticletitle{Compilers, principles, techniques}.
\newblock \bibinfo{journal}{\emph{Addison wesley}} \bibinfo{volume}{7}, \bibinfo{number}{8} (\bibinfo{year}{1986}), \bibinfo{pages}{9}.
\newblock


\bibitem[Al-Khanjari et~al\mbox{.}(2005)]%
        {article}
\bibfield{author}{\bibinfo{person}{Z. Al-Khanjari}, \bibinfo{person}{M. Woodward}, \bibinfo{person}{H. Ramadhan}, {and} \bibinfo{person}{N.~S. Kutti}.} \bibinfo{year}{2005}\natexlab{}.
\newblock \showarticletitle{Efficiency of Using Critical Slicing in Fault Localization}.
\newblock \bibinfo{journal}{\emph{Software Quality Journal}}  \bibinfo{volume}{13} (\bibinfo{date}{06} \bibinfo{year}{2005}), \bibinfo{pages}{129--153}.
\newblock
\urldef\tempurl%
\url{https://doi.org/10.1007/s11219-005-6214-x}
\showDOI{\tempurl}


\bibitem[{Alves} et~al\mbox{.}(2011)]%
        {6100114}
\bibfield{author}{\bibinfo{person}{E. {Alves}}, \bibinfo{person}{M. {Gligoric}}, \bibinfo{person}{V. {Jagannath}}, {and} \bibinfo{person}{M. {d'Amorim}}.} \bibinfo{year}{2011}\natexlab{}.
\newblock \showarticletitle{Fault-localization using dynamic slicing and change impact analysis}. In \bibinfo{booktitle}{\emph{2011 26th IEEE/ACM International Conference on Automated Software Engineering (ASE 2011)}}. \bibinfo{pages}{520--523}.
\newblock


\bibitem[B{\"o}hme and Roychoudhury(2014)]%
        {2014-ISSTA-Bohme}
\bibfield{author}{\bibinfo{person}{Marcel B{\"o}hme} {and} \bibinfo{person}{Abhik Roychoudhury}.} \bibinfo{year}{2014}\natexlab{}.
\newblock \showarticletitle{Corebench: Studying complexity of regression errors}. In \bibinfo{booktitle}{\emph{Proceedings of the 2014 International Symposium on Software Testing and Analysis}}. ACM, \bibinfo{pages}{105--115}.
\newblock


\bibitem[Cadar et~al\mbox{.}(2008)]%
        {2008-OSDI-Cadar}
\bibfield{author}{\bibinfo{person}{Cristian Cadar}, \bibinfo{person}{Daniel Dunbar}, {and} \bibinfo{person}{Dawson Engler}.} \bibinfo{year}{2008}\natexlab{}.
\newblock \showarticletitle{KLEE: Unassisted and Automatic Generation of High-coverage Tests for Complex Systems Programs}. In \bibinfo{booktitle}{\emph{Proceedings of the 8th USENIX Conference on Operating Systems Design and Implementation}} (San Diego, California) \emph{(\bibinfo{series}{OSDI'08})}. \bibinfo{publisher}{USENIX Association}, \bibinfo{address}{Berkeley, CA, USA}, \bibinfo{pages}{209--224}.
\newblock
\urldef\tempurl%
\url{http://dl.acm.org/citation.cfm?id=1855741.1855756}
\showURL{%
\tempurl}


\bibitem[Chen and Kim(2015)]%
        {6926857}
\bibfield{author}{\bibinfo{person}{Ning Chen} {and} \bibinfo{person}{Sunghun Kim}.} \bibinfo{year}{2015}\natexlab{}.
\newblock \showarticletitle{STAR: Stack Trace Based Automatic Crash Reproduction via Symbolic Execution}.
\newblock \bibinfo{journal}{\emph{IEEE Transactions on Software Engineering}} \bibinfo{volume}{41}, \bibinfo{number}{2} (\bibinfo{year}{2015}), \bibinfo{pages}{198--220}.
\newblock
\urldef\tempurl%
\url{https://doi.org/10.1109/TSE.2014.2363469}
\showDOI{\tempurl}


\bibitem[Cheng et~al\mbox{.}(2009)]%
        {2009-ISSTA-Cheng}
\bibfield{author}{\bibinfo{person}{Hong Cheng}, \bibinfo{person}{David Lo}, \bibinfo{person}{Yang Zhou}, \bibinfo{person}{Xiaoyin Wang}, {and} \bibinfo{person}{Xifeng Yan}.} \bibinfo{year}{2009}\natexlab{}.
\newblock \showarticletitle{Identifying Bug Signatures Using Discriminative Graph Mining}. In \bibinfo{booktitle}{\emph{Proceedings of the Eighteenth International Symposium on Software Testing and Analysis}} (Chicago, IL, USA) \emph{(\bibinfo{series}{ISSTA '09})}. \bibinfo{publisher}{ACM}, \bibinfo{address}{New York, NY, USA}, \bibinfo{pages}{141--152}.
\newblock
\showISBNx{978-1-60558-338-9}
\urldef\tempurl%
\url{https://doi.org/10.1145/1572272.1572290}
\showDOI{\tempurl}


\bibitem[Das et~al\mbox{.}(2002)]%
        {ESP}
\bibfield{author}{\bibinfo{person}{Manuvir Das}, \bibinfo{person}{Sorin Lerner}, {and} \bibinfo{person}{Mark Seigle}.} \bibinfo{year}{2002}\natexlab{}.
\newblock \showarticletitle{ESP: Path-Sensitive Program Verification in Polynomial Time}. In \bibinfo{booktitle}{\emph{Proceedings of the ACM SIGPLAN 2002 Conference on Programming Language Design and Implementation}} (Berlin, Germany) \emph{(\bibinfo{series}{PLDI '02})}. \bibinfo{publisher}{Association for Computing Machinery}, \bibinfo{address}{New York, NY, USA}, \bibinfo{pages}{57–68}.
\newblock
\showISBNx{1581134630}
\urldef\tempurl%
\url{https://doi.org/10.1145/512529.512538}
\showDOI{\tempurl}


\bibitem[DeMillo et~al\mbox{.}(1996)]%
        {DeMillo:1996:ISSTA}
\bibfield{author}{\bibinfo{person}{Richard~A. DeMillo}, \bibinfo{person}{Hsin Pan}, {and} \bibinfo{person}{Eugene~H. Spafford}.} \bibinfo{year}{1996}\natexlab{}.
\newblock \showarticletitle{Critical Slicing for Software Fault Localization}. In \bibinfo{booktitle}{\emph{Proceedings of the 1996 ACM SIGSOFT International Symposium on Software Testing and Analysis}} (San Diego, California, USA) \emph{(\bibinfo{series}{ISSTA '96})}. \bibinfo{publisher}{ACM}, \bibinfo{address}{New York, NY, USA}, \bibinfo{pages}{121--134}.
\newblock
\showISBNx{0-89791-787-1}
\urldef\tempurl%
\url{https://doi.org/10.1145/229000.226310}
\showDOI{\tempurl}


\bibitem[Ernst et~al\mbox{.}(1999)]%
        {1999-ICSE-Ernst}
\bibfield{author}{\bibinfo{person}{Michael~D. Ernst}, \bibinfo{person}{Jake Cockrell}, \bibinfo{person}{William~G. Griswold}, {and} \bibinfo{person}{David Notkin}.} \bibinfo{year}{1999}\natexlab{}.
\newblock \showarticletitle{Dynamically Discovering Likely Program Invariants to Support Program Evolution}. In \bibinfo{booktitle}{\emph{Proceedings of the 21st International Conference on Software Engineering}} (Los Angeles, California, USA) \emph{(\bibinfo{series}{ICSE '99})}. \bibinfo{publisher}{ACM}, \bibinfo{address}{New York, NY, USA}, \bibinfo{pages}{213--224}.
\newblock
\showISBNx{1-58113-074-0}
\urldef\tempurl%
\url{https://doi.org/10.1145/302405.302467}
\showDOI{\tempurl}


\bibitem[Golagha et~al\mbox{.}(2018)]%
        {8449433}
\bibfield{author}{\bibinfo{person}{Mojdeh Golagha}, \bibinfo{person}{Abu~Mohammed Raisuddin}, \bibinfo{person}{Lennart Mittag}, \bibinfo{person}{Dominik Hellhake}, {and} \bibinfo{person}{Alexander Pretschner}.} \bibinfo{year}{2018}\natexlab{}.
\newblock \showarticletitle{Aletheia: A Failure Diagnosis Toolchain}. In \bibinfo{booktitle}{\emph{2018 IEEE/ACM 40th International Conference on Software Engineering: Companion (ICSE-Companion)}}. \bibinfo{pages}{13--16}.
\newblock


\bibitem[Gyim\'{o}thy et~al\mbox{.}(1999)]%
        {Gyimothy:1999:FSE}
\bibfield{author}{\bibinfo{person}{Tibor Gyim\'{o}thy}, \bibinfo{person}{\'{A}rp\'{a}d Besz{\'e}des}, {and} \bibinfo{person}{Ist\'{a}n Forg\'{a}cs}.} \bibinfo{year}{1999}\natexlab{}.
\newblock \showarticletitle{An Efficient Relevant Slicing Method for Debugging}. In \bibinfo{booktitle}{\emph{Proceedings of the 7th European Software Engineering Conference Held Jointly with the 7th ACM SIGSOFT International Symposium on Foundations of Software Engineering}} (Toulouse, France) \emph{(\bibinfo{series}{ESEC/FSE-7})}. \bibinfo{publisher}{Springer-Verlag}, \bibinfo{address}{Berlin, Heidelberg}, \bibinfo{pages}{303--321}.
\newblock
\showISBNx{3-540-66538-2}
\urldef\tempurl%
\url{http://dl.acm.org/citation.cfm?id=318773.319248}
\showURL{%
\tempurl}


\bibitem[Helin(2016)]%
        {helin2016radamsa}
\bibfield{author}{\bibinfo{person}{Aki Helin}.} \bibinfo{year}{2016}\natexlab{}.
\newblock \showarticletitle{Radamsa}.
\newblock \bibinfo{journal}{\emph{URL: https://gitlab. com/akihe/radamsa}} (\bibinfo{year}{2016}).
\newblock


\bibitem[Hsu et~al\mbox{.}(2008)]%
        {2008-ASE-Hwa-You}
\bibfield{author}{\bibinfo{person}{Hwa-You Hsu}, \bibinfo{person}{J.~A. Jones}, {and} \bibinfo{person}{A. Orso}.} \bibinfo{year}{2008}\natexlab{}.
\newblock \showarticletitle{Rapid: Identifying Bug Signatures to Support Debugging Activities}. In \bibinfo{booktitle}{\emph{Proceedings of the 2008 23rd IEEE/ACM International Conference on Automated Software Engineering}} \emph{(\bibinfo{series}{ASE '08})}. \bibinfo{publisher}{IEEE Computer Society}, \bibinfo{address}{Washington, DC, USA}, \bibinfo{pages}{439--442}.
\newblock
\showISBNx{978-1-4244-2187-9}
\urldef\tempurl%
\url{https://doi.org/10.1109/ASE.2008.68}
\showDOI{\tempurl}


\bibitem[Jones and Harrold(2005)]%
        {Jones:2005:ASE}
\bibfield{author}{\bibinfo{person}{James~A. Jones} {and} \bibinfo{person}{Mary~Jean Harrold}.} \bibinfo{year}{2005}\natexlab{}.
\newblock \showarticletitle{Empirical Evaluation of the Tarantula Automatic Fault-localization Technique}. In \bibinfo{booktitle}{\emph{Proceedings of the 20th IEEE/ACM International Conference on Automated Software Engineering}} (Long Beach, CA, USA) \emph{(\bibinfo{series}{ASE '05})}. \bibinfo{publisher}{ACM}, \bibinfo{address}{New York, NY, USA}, \bibinfo{pages}{273--282}.
\newblock
\showISBNx{1-58113-993-4}
\urldef\tempurl%
\url{https://doi.org/10.1145/1101908.1101949}
\showDOI{\tempurl}


\bibitem[{Jones} et~al\mbox{.}(2002)]%
        {Jones:2002:ICSE}
\bibfield{author}{\bibinfo{person}{J.~A. {Jones}}, \bibinfo{person}{M.~J. {Harrold}}, {and} \bibinfo{person}{J. {Stasko}}.} \bibinfo{year}{2002}\natexlab{}.
\newblock \showarticletitle{Visualization of test information to assist fault localization}. In \bibinfo{booktitle}{\emph{Proceedings of the 24th International Conference on Software Engineering. ICSE 2002}}. \bibinfo{pages}{467--477}.
\newblock
\urldef\tempurl%
\url{https://doi.org/10.1145/581396.581397}
\showDOI{\tempurl}


\bibitem[Jones et~al\mbox{.}(2002)]%
        {2002-ICES-Jones}
\bibfield{author}{\bibinfo{person}{James~A. Jones}, \bibinfo{person}{Mary~Jean Harrold}, {and} \bibinfo{person}{John Stasko}.} \bibinfo{year}{2002}\natexlab{}.
\newblock \showarticletitle{Visualization of Test Information to Assist Fault Localization}. In \bibinfo{booktitle}{\emph{Proceedings of the 24th International Conference on Software Engineering}} (Orlando, Florida) \emph{(\bibinfo{series}{ICSE '02})}. \bibinfo{publisher}{Association for Computing Machinery}, \bibinfo{address}{New York, NY, USA}, \bibinfo{pages}{467–477}.
\newblock
\showISBNx{158113472X}
\urldef\tempurl%
\url{https://doi.org/10.1145/581339.581397}
\showDOI{\tempurl}


\bibitem[Ju et~al\mbox{.}(2013)]%
        {HSFal405}
\bibfield{author}{\bibinfo{person}{Xiaolin Ju}, \bibinfo{person}{Shujuan Jiang}, \bibinfo{person}{Xiang Chen}, \bibinfo{person}{Xingya Wang}, \bibinfo{person}{Zhang Yanmei}, {and} \bibinfo{person}{Heling Cao}.} \bibinfo{year}{2013}\natexlab{}.
\newblock \showarticletitle{HSFal: Effective Fault Localization using Hybrid Spectrum of Full Slices and Execution Slices}.
\newblock \bibinfo{journal}{\emph{Journal of Systems and Software}}  \bibinfo{volume}{90} (\bibinfo{date}{01} \bibinfo{year}{2013}).
\newblock
\urldef\tempurl%
\url{https://doi.org/10.1016/j.jss.2013.11.1109}
\showDOI{\tempurl}


\bibitem[Kallingal~Joshy et~al\mbox{.}(2021)]%
        {kallingal2021validating}
\bibfield{author}{\bibinfo{person}{Ashwin Kallingal~Joshy}, \bibinfo{person}{Xueyuan Chen}, \bibinfo{person}{Benjamin Steenhoek}, {and} \bibinfo{person}{Wei Le}.} \bibinfo{year}{2021}\natexlab{}.
\newblock \showarticletitle{Validating static warnings via testing code fragments}. In \bibinfo{booktitle}{\emph{Proceedings of the 30th ACM SIGSOFT International Symposium on Software Testing and Analysis}}. \bibinfo{pages}{540--552}.
\newblock


\bibitem[Kasikci et~al\mbox{.}(2015)]%
        {2015-SOSP-Kasikci}
\bibfield{author}{\bibinfo{person}{Baris Kasikci}, \bibinfo{person}{Benjamin Schubert}, \bibinfo{person}{Cristiano Pereira}, \bibinfo{person}{Gilles Pokam}, {and} \bibinfo{person}{George Candea}.} \bibinfo{year}{2015}\natexlab{}.
\newblock \showarticletitle{Failure Sketching: A Technique for Automated Root Cause Diagnosis of in-Production Failures}. In \bibinfo{booktitle}{\emph{Proceedings of the 25th Symposium on Operating Systems Principles}} (Monterey, California) \emph{(\bibinfo{series}{SOSP '15})}. \bibinfo{publisher}{Association for Computing Machinery}, \bibinfo{address}{New York, NY, USA}, \bibinfo{pages}{344–360}.
\newblock
\showISBNx{9781450338349}
\urldef\tempurl%
\url{https://doi.org/10.1145/2815400.2815412}
\showDOI{\tempurl}


\bibitem[Landi(1992)]%
        {landi1992undecidability}
\bibfield{author}{\bibinfo{person}{William Landi}.} \bibinfo{year}{1992}\natexlab{}.
\newblock \showarticletitle{Undecidability of static analysis}.
\newblock \bibinfo{journal}{\emph{ACM Letters on Programming Languages and Systems (LOPLAS)}} \bibinfo{volume}{1}, \bibinfo{number}{4} (\bibinfo{year}{1992}), \bibinfo{pages}{323--337}.
\newblock


\bibitem[Le and Soffa(2008)]%
        {le2008marple}
\bibfield{author}{\bibinfo{person}{Wei Le} {and} \bibinfo{person}{Mary~Lou Soffa}.} \bibinfo{year}{2008}\natexlab{}.
\newblock \showarticletitle{Marple: a demand-driven path-sensitive buffer overflow detector}. In \bibinfo{booktitle}{\emph{Proceedings of the 16th ACM SIGSOFT International Symposium on Foundations of software engineering}}. ACM, \bibinfo{pages}{272--282}.
\newblock


\bibitem[Le~Goues et~al\mbox{.}(2015)]%
        {2015-IEE-Goues}
\bibfield{author}{\bibinfo{person}{Claire Le~Goues}, \bibinfo{person}{Neal Holtschulte}, \bibinfo{person}{Edward~K. Smith}, \bibinfo{person}{Yuriy Brun}, \bibinfo{person}{Premkumar Devanbu}, \bibinfo{person}{Stephanie Forrest}, {and} \bibinfo{person}{Westley Weimer}.} \bibinfo{year}{2015}\natexlab{}.
\newblock \showarticletitle{The ManyBugs and IntroClass Benchmarks for Automated Repair of C Programs}.
\newblock \bibinfo{journal}{\emph{IEEE Transactions on Software Engineering}} \bibinfo{volume}{41}, \bibinfo{number}{12} (\bibinfo{year}{2015}), \bibinfo{pages}{1236--1256}.
\newblock
\urldef\tempurl%
\url{https://doi.org/10.1109/TSE.2015.2454513}
\showDOI{\tempurl}


\bibitem[{Li} et~al\mbox{.}(2017)]%
        {2017-IEEE-Li}
\bibfield{author}{\bibinfo{person}{F. {Li}}, \bibinfo{person}{Z. {Li}}, \bibinfo{person}{W. {Huo}}, {and} \bibinfo{person}{X. {Feng}}.} \bibinfo{year}{2017}\natexlab{}.
\newblock \showarticletitle{Locating Software Faults Based on Minimum Debugging Frontier Set}.
\newblock \bibinfo{journal}{\emph{IEEE Transactions on Software Engineering}} \bibinfo{volume}{43}, \bibinfo{number}{8} (\bibinfo{year}{2017}), \bibinfo{pages}{760--776}.
\newblock


\bibitem[{Lo} et~al\mbox{.}(2011)]%
        {2011-QRSC-Lo}
\bibfield{author}{\bibinfo{person}{D. {Lo}}, \bibinfo{person}{H. {Cheng}}, {and} \bibinfo{person}{X. {Wang}}.} \bibinfo{year}{2011}\natexlab{}.
\newblock \showarticletitle{Bug Signature Minimization and Fusion}. In \bibinfo{booktitle}{\emph{2011 IEEE 13th International Symposium on High-Assurance Systems Engineering}}. \bibinfo{pages}{340--347}.
\newblock


\bibitem[Lu et~al\mbox{.}(2005a)]%
        {2005-Workshop-Lu}
\bibfield{author}{\bibinfo{person}{Shan Lu}, \bibinfo{person}{Zhenmin Li}, \bibinfo{person}{Feng Qin}, \bibinfo{person}{Lin Tan}, \bibinfo{person}{Pin Zhou}, {and} \bibinfo{person}{Yuanyuan Zhou}.} \bibinfo{year}{2005}\natexlab{a}.
\newblock \showarticletitle{Bugbench: Benchmarks for evaluating bug detection tools}. In \bibinfo{booktitle}{\emph{Workshop on the evaluation of software defect detection tools}}, Vol.~\bibinfo{volume}{5}.
\newblock


\bibitem[Lu et~al\mbox{.}(2005b)]%
        {lu2005bugbench}
\bibfield{author}{\bibinfo{person}{Shan Lu}, \bibinfo{person}{Zhenmin Li}, \bibinfo{person}{Feng Qin}, \bibinfo{person}{Lin Tan}, \bibinfo{person}{Pin Zhou}, {and} \bibinfo{person}{Yuanyuan Zhou}.} \bibinfo{year}{2005}\natexlab{b}.
\newblock \showarticletitle{Bugbench: Benchmarks for evaluating bug detection tools}. In \bibinfo{booktitle}{\emph{Workshop on the evaluation of software defect detection tools}}, Vol.~\bibinfo{volume}{5}.
\newblock


\bibitem[Man{\`{e}}s et~al\mbox{.}(2018)]%
        {DBLP:journals/corr/abs-1812-00140}
\bibfield{author}{\bibinfo{person}{Valentin J.~M. Man{\`{e}}s}, \bibinfo{person}{HyungSeok Han}, \bibinfo{person}{Choongwoo Han}, \bibinfo{person}{Sang~Kil Cha}, \bibinfo{person}{Manuel Egele}, \bibinfo{person}{Edward~J. Schwartz}, {and} \bibinfo{person}{Maverick Woo}.} \bibinfo{year}{2018}\natexlab{}.
\newblock \showarticletitle{Fuzzing: Art, Science, and Engineering}.
\newblock \bibinfo{journal}{\emph{CoRR}}  \bibinfo{volume}{abs/1812.00140} (\bibinfo{year}{2018}).
\newblock
\showeprint[arxiv]{1812.00140}
\urldef\tempurl%
\url{http://arxiv.org/abs/1812.00140}
\showURL{%
\tempurl}


\bibitem[Manevich et~al\mbox{.}(2004)]%
        {2004-FSE-Manevich}
\bibfield{author}{\bibinfo{person}{Roman Manevich}, \bibinfo{person}{Manu Sridharan}, \bibinfo{person}{Stephen Adams}, \bibinfo{person}{Manuvir Das}, {and} \bibinfo{person}{Zhe Yang}.} \bibinfo{year}{2004}\natexlab{}.
\newblock \showarticletitle{PSE: explaining program failures via postmortem static analysis}. In \bibinfo{booktitle}{\emph{ACM SIGSOFT Software Engineering Notes}}, Vol.~\bibinfo{volume}{29}. ACM, \bibinfo{pages}{63--72}.
\newblock


\bibitem[McCabe(1976)]%
        {1976-IEE-McCabe}
\bibfield{author}{\bibinfo{person}{T.J. McCabe}.} \bibinfo{year}{1976}\natexlab{}.
\newblock \showarticletitle{A Complexity Measure}.
\newblock \bibinfo{journal}{\emph{IEEE Transactions on Software Engineering}} \bibinfo{volume}{SE-2}, \bibinfo{number}{4} (\bibinfo{year}{1976}), \bibinfo{pages}{308--320}.
\newblock
\urldef\tempurl%
\url{https://doi.org/10.1109/TSE.1976.233837}
\showDOI{\tempurl}


\bibitem[Mechtaev et~al\mbox{.}(2016)]%
        {angelix}
\bibfield{author}{\bibinfo{person}{Sergey Mechtaev}, \bibinfo{person}{Jooyong Yi}, {and} \bibinfo{person}{Abhik Roychoudhury}.} \bibinfo{year}{2016}\natexlab{}.
\newblock \showarticletitle{Angelix: Scalable Multiline Program Patch Synthesis via Symbolic Analysis}. In \bibinfo{booktitle}{\emph{Proceedings of the 38th International Conference on Software Engineering}} (Austin, Texas) \emph{(\bibinfo{series}{ICSE '16})}. \bibinfo{publisher}{ACM}, \bibinfo{address}{New York, NY, USA}, \bibinfo{pages}{691--701}.
\newblock
\showISBNx{978-1-4503-3900-1}
\urldef\tempurl%
\url{https://doi.org/10.1145/2884781.2884807}
\showDOI{\tempurl}


\bibitem[Nayrolles et~al\mbox{.}(2015)]%
        {7081820}
\bibfield{author}{\bibinfo{person}{Mathieu Nayrolles}, \bibinfo{person}{Abdelwahab Hamou-Lhadj}, \bibinfo{person}{Sofiène Tahar}, {and} \bibinfo{person}{Alf Larsson}.} \bibinfo{year}{2015}\natexlab{}.
\newblock \showarticletitle{JCHARMING: A bug reproduction approach using crash traces and directed model checking}. In \bibinfo{booktitle}{\emph{2015 IEEE 22nd International Conference on Software Analysis, Evolution, and Reengineering (SANER)}}. \bibinfo{pages}{101--110}.
\newblock
\urldef\tempurl%
\url{https://doi.org/10.1109/SANER.2015.7081820}
\showDOI{\tempurl}


\bibitem[Pan et~al\mbox{.}(2020)]%
        {2020-SOFL-Pan}
\bibfield{author}{\bibinfo{person}{Baoyi Pan}, \bibinfo{person}{Ting Shu}, \bibinfo{person}{Jinsong Xia}, \bibinfo{person}{Zuohua Ding}, {and} \bibinfo{person}{Mingyue Jiang}.} \bibinfo{year}{2020}\natexlab{}.
\newblock \showarticletitle{A Fault Localization Method Based on Dynamic Failed Execution Blocks}. In \bibinfo{booktitle}{\emph{Structured Object-Oriented Formal Language and Method}}, \bibfield{editor}{\bibinfo{person}{Huaikou Miao}, \bibinfo{person}{Cong Tian}, \bibinfo{person}{Shaoying Liu}, {and} \bibinfo{person}{Zhenhua Duan}} (Eds.). \bibinfo{publisher}{Springer International Publishing}, \bibinfo{address}{Cham}, \bibinfo{pages}{315--327}.
\newblock
\showISBNx{978-3-030-41418-4}


\bibitem[P{\'e}rez et~al\mbox{.}(2019)]%
        {Prez2019AutomaticTO}
\bibfield{author}{\bibinfo{person}{Sergio P{\'e}rez}, \bibinfo{person}{Josep Silva}, {and} \bibinfo{person}{Salvador Tamarit}.} \bibinfo{year}{2019}\natexlab{}.
\newblock \showarticletitle{Automatic Testing of Program Slicers}.
\newblock \bibinfo{journal}{\emph{Sci. Program.}}  \bibinfo{volume}{2019} (\bibinfo{year}{2019}), \bibinfo{pages}{4108652:1--4108652:15}.
\newblock


\bibitem[{Renieres} and {Reiss}(2003)]%
        {2003-IEEE-Renieres}
\bibfield{author}{\bibinfo{person}{M. {Renieres}} {and} \bibinfo{person}{S.~P. {Reiss}}.} \bibinfo{year}{2003}\natexlab{}.
\newblock \showarticletitle{Fault localization with nearest neighbor queries}. In \bibinfo{booktitle}{\emph{18th IEEE International Conference on Automated Software Engineering, 2003. Proceedings.}} \bibinfo{pages}{30--39}.
\newblock


\bibitem[Soltani et~al\mbox{.}(2017)]%
        {7985663}
\bibfield{author}{\bibinfo{person}{Mozhan Soltani}, \bibinfo{person}{Annibale Panichella}, {and} \bibinfo{person}{Arie van Deursen}.} \bibinfo{year}{2017}\natexlab{}.
\newblock \showarticletitle{A Guided Genetic Algorithm for Automated Crash Reproduction}. In \bibinfo{booktitle}{\emph{2017 IEEE/ACM 39th International Conference on Software Engineering (ICSE)}}. \bibinfo{pages}{209--220}.
\newblock
\urldef\tempurl%
\url{https://doi.org/10.1109/ICSE.2017.27}
\showDOI{\tempurl}


\bibitem[Sun and Khoo(2013)]%
        {2013-FSE-Sun}
\bibfield{author}{\bibinfo{person}{Chengnian Sun} {and} \bibinfo{person}{Siau-Cheng Khoo}.} \bibinfo{year}{2013}\natexlab{}.
\newblock \showarticletitle{Mining Succinct Predicated Bug Signatures}. In \bibinfo{booktitle}{\emph{Proceedings of the 2013 9th Joint Meeting on Foundations of Software Engineering}} (Saint Petersburg, Russia) \emph{(\bibinfo{series}{ESEC/FSE 2013})}. \bibinfo{publisher}{Association for Computing Machinery}, \bibinfo{address}{New York, NY, USA}, \bibinfo{pages}{576–586}.
\newblock
\showISBNx{9781450322379}
\urldef\tempurl%
\url{https://doi.org/10.1145/2491411.2491449}
\showDOI{\tempurl}


\bibitem[{Wang} and {Huang}(2016)]%
        {2016-QRSC-Wang}
\bibfield{author}{\bibinfo{person}{Y. {Wang}} {and} \bibinfo{person}{Z. {Huang}}.} \bibinfo{year}{2016}\natexlab{}.
\newblock \showarticletitle{Weighted Control Flow Subgraph to Support Debugging Activities}. In \bibinfo{booktitle}{\emph{2016 IEEE International Conference on Software Quality, Reliability and Security Companion (QRS-C)}}. \bibinfo{pages}{131--134}.
\newblock


\bibitem[Weiser(1981)]%
        {weiser1981program}
\bibfield{author}{\bibinfo{person}{Mark Weiser}.} \bibinfo{year}{1981}\natexlab{}.
\newblock \showarticletitle{Program slicing}. In \bibinfo{booktitle}{\emph{Proceedings of the 5th international conference on Software engineering}}. IEEE Press, \bibinfo{pages}{439--449}.
\newblock


\bibitem[{Wong} et~al\mbox{.}(2014)]%
        {2014-IEEE-Wong}
\bibfield{author}{\bibinfo{person}{W.~E. {Wong}}, \bibinfo{person}{V. {Debroy}}, \bibinfo{person}{R. {Gao}}, {and} \bibinfo{person}{Y. {Li}}.} \bibinfo{year}{2014}\natexlab{}.
\newblock \showarticletitle{The DStar Method for Effective Software Fault Localization}.
\newblock \bibinfo{journal}{\emph{IEEE Transactions on Reliability}} \bibinfo{volume}{63}, \bibinfo{number}{1} (\bibinfo{year}{2014}), \bibinfo{pages}{290--308}.
\newblock


\bibitem[{Wong} et~al\mbox{.}(2016)]%
        {2016-IEEE-Wong}
\bibfield{author}{\bibinfo{person}{W.~E. {Wong}}, \bibinfo{person}{R. {Gao}}, \bibinfo{person}{Y. {Li}}, \bibinfo{person}{R. {Abreu}}, {and} \bibinfo{person}{F. {Wotawa}}.} \bibinfo{year}{2016}\natexlab{}.
\newblock \showarticletitle{A Survey on Software Fault Localization}.
\newblock \bibinfo{journal}{\emph{IEEE Transactions on Software Engineering}} \bibinfo{volume}{42}, \bibinfo{number}{8} (\bibinfo{year}{2016}), \bibinfo{pages}{707--740}.
\newblock


\bibitem[{Wotawa}(2010)]%
        {5562955}
\bibfield{author}{\bibinfo{person}{F. {Wotawa}}.} \bibinfo{year}{2010}\natexlab{}.
\newblock \showarticletitle{Fault Localization Based on Dynamic Slicing and Hitting-Set Computation}. In \bibinfo{booktitle}{\emph{2010 10th International Conference on Quality Software}}. \bibinfo{pages}{161--170}.
\newblock


\bibitem[Xie and Aiken(2007)]%
        {xie2007saturn}
\bibfield{author}{\bibinfo{person}{Yichen Xie} {and} \bibinfo{person}{Alex Aiken}.} \bibinfo{year}{2007}\natexlab{}.
\newblock \showarticletitle{Saturn: A scalable framework for error detection using boolean satisfiability}.
\newblock \bibinfo{journal}{\emph{ACM Transactions on Programming Languages and Systems (TOPLAS)}} \bibinfo{volume}{29}, \bibinfo{number}{3} (\bibinfo{year}{2007}), \bibinfo{pages}{16--es}.
\newblock


\bibitem[Zalewski(2014)]%
        {zalewski2014american}
\bibfield{author}{\bibinfo{person}{Michal Zalewski}.} \bibinfo{year}{2014}\natexlab{}.
\newblock \showarticletitle{American fuzzy lop (2017)}.
\newblock \bibinfo{journal}{\emph{URL http://lcamtuf. coredump. cx/afl}}  \bibinfo{volume}{14} (\bibinfo{year}{2014}), \bibinfo{pages}{28}.
\newblock


\bibitem[Zhang et~al\mbox{.}(2006)]%
        {Zhang:2006:PLDI}
\bibfield{author}{\bibinfo{person}{Xiangyu Zhang}, \bibinfo{person}{Neelam Gupta}, {and} \bibinfo{person}{Rajiv Gupta}.} \bibinfo{year}{2006}\natexlab{}.
\newblock \showarticletitle{Pruning Dynamic Slices with Confidence}. In \bibinfo{booktitle}{\emph{Proceedings of the 27th ACM SIGPLAN Conference on Programming Language Design and Implementation}} (Ottawa, Ontario, Canada) \emph{(\bibinfo{series}{PLDI '06})}. \bibinfo{publisher}{ACM}, \bibinfo{address}{New York, NY, USA}, \bibinfo{pages}{169--180}.
\newblock
\showISBNx{1-59593-320-4}
\urldef\tempurl%
\url{https://doi.org/10.1145/1133981.1134002}
\showDOI{\tempurl}


\bibitem[Zhao et~al\mbox{.}(2019)]%
        {8811942}
\bibfield{author}{\bibinfo{person}{Yu Zhao}, \bibinfo{person}{Tingting Yu}, \bibinfo{person}{Ting Su}, \bibinfo{person}{Yang Liu}, \bibinfo{person}{Wei Zheng}, \bibinfo{person}{Jingzhi Zhang}, {and} \bibinfo{person}{William G.J.~Halfond}.} \bibinfo{year}{2019}\natexlab{}.
\newblock \showarticletitle{ReCDroid: Automatically Reproducing Android Application Crashes from Bug Reports}. In \bibinfo{booktitle}{\emph{2019 IEEE/ACM 41st International Conference on Software Engineering (ICSE)}}. \bibinfo{pages}{128--139}.
\newblock
\urldef\tempurl%
\url{https://doi.org/10.1109/ICSE.2019.00030}
\showDOI{\tempurl}


\bibitem[Zitser et~al\mbox{.}(2004)]%
        {zitser2004testing}
\bibfield{author}{\bibinfo{person}{Misha Zitser}, \bibinfo{person}{Richard Lippmann}, {and} \bibinfo{person}{Tim Leek}.} \bibinfo{year}{2004}\natexlab{}.
\newblock \showarticletitle{Testing static analysis tools using exploitable buffer overflows from open source code}. In \bibinfo{booktitle}{\emph{Proceedings of the 12th ACM SIGSOFT twelfth international symposium on Foundations of software engineering}}. \bibinfo{pages}{97--106}.
\newblock


\bibitem[Zuo et~al\mbox{.}(2014)]%
        {2014-ISSTA-Zuo}
\bibfield{author}{\bibinfo{person}{Zhiqiang Zuo}, \bibinfo{person}{Siau-Cheng Khoo}, {and} \bibinfo{person}{Chengnian Sun}.} \bibinfo{year}{2014}\natexlab{}.
\newblock \showarticletitle{Efficient Predicated Bug Signature Mining via Hierarchical Instrumentation}. In \bibinfo{booktitle}{\emph{Proceedings of the 2014 International Symposium on Software Testing and Analysis}} (San Jose, CA, USA) \emph{(\bibinfo{series}{ISSTA 2014})}. \bibinfo{publisher}{ACM}, \bibinfo{address}{New York, NY, USA}, \bibinfo{pages}{215--224}.
\newblock
\showISBNx{978-1-4503-2645-2}
\urldef\tempurl%
\url{https://doi.org/10.1145/2610384.2610400}
\showDOI{\tempurl}


\end{thebibliography}

\end{document}